\documentclass[sigconf, nonacm, pdfa]{acmart}

\hypersetup{pdfapart=2,pdfaconformance=B,pdflang=en-US}
\makeatletter
\renewcommand*{\hyxmp@xmptpg@schema}{}
\makeatother
\immediate\pdfobj stream attr{/N 3} file{sRGB.icc}
\edef\vldbsrgbprofile{\the\pdflastobj}
\immediate\pdfobj{<<
  /Type /OutputIntent
  /S /GTS_PDFA1
  /OutputConditionIdentifier (sRGB IEC61966-2.1)
  /Info (sRGB IEC61966-2.1)
  /RegistryName (http://www.color.org)
  /DestOutputProfile \vldbsrgbprofile\space 0 R
>>}
\pdfcatalog{/OutputIntents [\the\pdflastobj\space 0 R]}

\usepackage{pvldb}%

\renewcommand\vldbvolume{20}
\renewcommand\vldbissue{3}
\renewcommand\vldbyear{2026}
\renewcommand\vldbdoi{XX.XX/XXX.XX}
\renewcommand\vldbpages{XXX-XXX}
\renewcommand\vldbpagestyle{empty}
\renewcommand\vldbavailabilityurl{https://github.com/fukien/contention-aware-spgemm}

\usepackage{xspace}
\newcommand{\myapp}{AMS\xspace}
\newcommand{\myappfull}{adaptive morsel scheduling\xspace}

\newcommand{\mysch}{CALB\xspace}
\newcommand{\myschfull}{contention-aware load balancing\xspace}
\newcommand{\myschfullcap}{Contention-Aware Load Balancing\xspace}

\newcommand{\mydaha}{SMTact\xspace}
\newcommand{\mydahafull}{data-conscious simultaneous multithreading activator\xspace}
\newcommand{\mydahafullcap}{Data-Conscious Simultaneous Multithreading Activator\xspace}

\newcommand{\myaccum}{HYB\xspace}
\newcommand{\myaccumfull}{hybrid accumulation mechanism\xspace}
\newcommand{\myaccumfullcap}{Hybrid Accumulation Mechanism\xspace}

\usepackage{verbatim}

\usepackage{tikz}
\usetikzlibrary{matrix, arrows.meta, positioning, calc, decorations.pathreplacing}

\usepackage{listings}
\usepackage{xcolor}

\usepackage{algorithm}
\usepackage{algpseudocode}

\usepackage{subcaption}

\usepackage{wrapfig}

\usepackage{bm}
\usepackage[table]{xcolor}
\usepackage{pifont}
\newcommand{\cmark}{\ding{51}}
\newcommand{\xmark}{\ding{55}}

\usepackage{tabularx}
\usepackage{multirow}

\setcitestyle{numbers,sort&compress}

\usepackage{enumitem}

\begin{document}
\title{Taming Multi-Dimensional Skew in Sparse Matrix Multiplication with Contention-Aware Scheduling [Technical Report]}

\author{Wentao Huang}
\affiliation{%
	\institution{National University of Singapore} 
	\country{}
}
\email{huangwentao@u.nus.edu}

\author{Qiming Huang}
\affiliation{
	\institution{National University of Singapore}
	\country{}
}
\email{huang\_qiming@u.nus.edu}

\author{Yunhong Ji}
\affiliation{
	\institution{Renmin University of China}
	\country{}
}
\email{jiyunhong@ruc.edu.cn}

\author{Mian Lu}
\affiliation{
	\institution{Phancy Group Co., Ltd.}
	\country{}
}
\email{lumian@4paradigm.com}

\author{Yuqiang Chen}
\affiliation{
	\institution{Phancy Group Co., Ltd.}
	\country{}
}
\email{chenyuqiang@4paradigm.com}

\author{Bingsheng He}
\affiliation{%
	\institution{National University of Singapore}
	\country{}
}
\email{dcsheb@nus.edu.sg}

\author{Kian-Lee Tan}
\affiliation{%
	\institution{National University of Singapore}
	\country{}
}
\email{dcstankl@nus.edu.sg}

\begin{abstract}
	Sparse general matrix-matrix multiplication (SpGEMM) is a key kernel in graph analytics, relational queries, and machine learning.
	Yet, its performance remains a significant challenge, due to the inherent sparsity and, more critically, the prevalent skew patterns often found in real-world workloads.
	This paper investigates this problem and demonstrates that an unbalanced workload distribution is the main culprit, harming performance through two co-existing root causes: first, \emph{multi-dimensional skew}, where imbalance across computation cost, output density, and data locality cannot be captured by any single balancing metric; and second, \emph{cache contention}, where dense substructures induce severe on-chip interference.

	To that end, we propose \myappfull{} (\myapp{}), a dynamic scheduling framework that decouples skew dimensions via fixed-result-size morsels and integrates \myschfull{} (\mysch{}) and \mydahafull{} (\mydaha{}) to alleviate cache-level contention, alongside a \myaccumfull{} (\myaccum{}) that adaptively selects the best accumulator per morsel.
	Evaluations across matrix squaring, relational queries, graph analytics, and other sparse matrix kernels show that \myapp{} significantly outperforms state-of-the-art methods on skewed workloads while maintaining comparable performance on non-skewed ones.
	These gains, along with improved bandwidth utilization, remain valid across a wide range of workloads and hardware systems.
\end{abstract}

\maketitle

\vldbtopmatter

\section{Introduction}
\label{sec:intro}

\begin{figure}
	\centering

	\includegraphics[width=\linewidth
	]{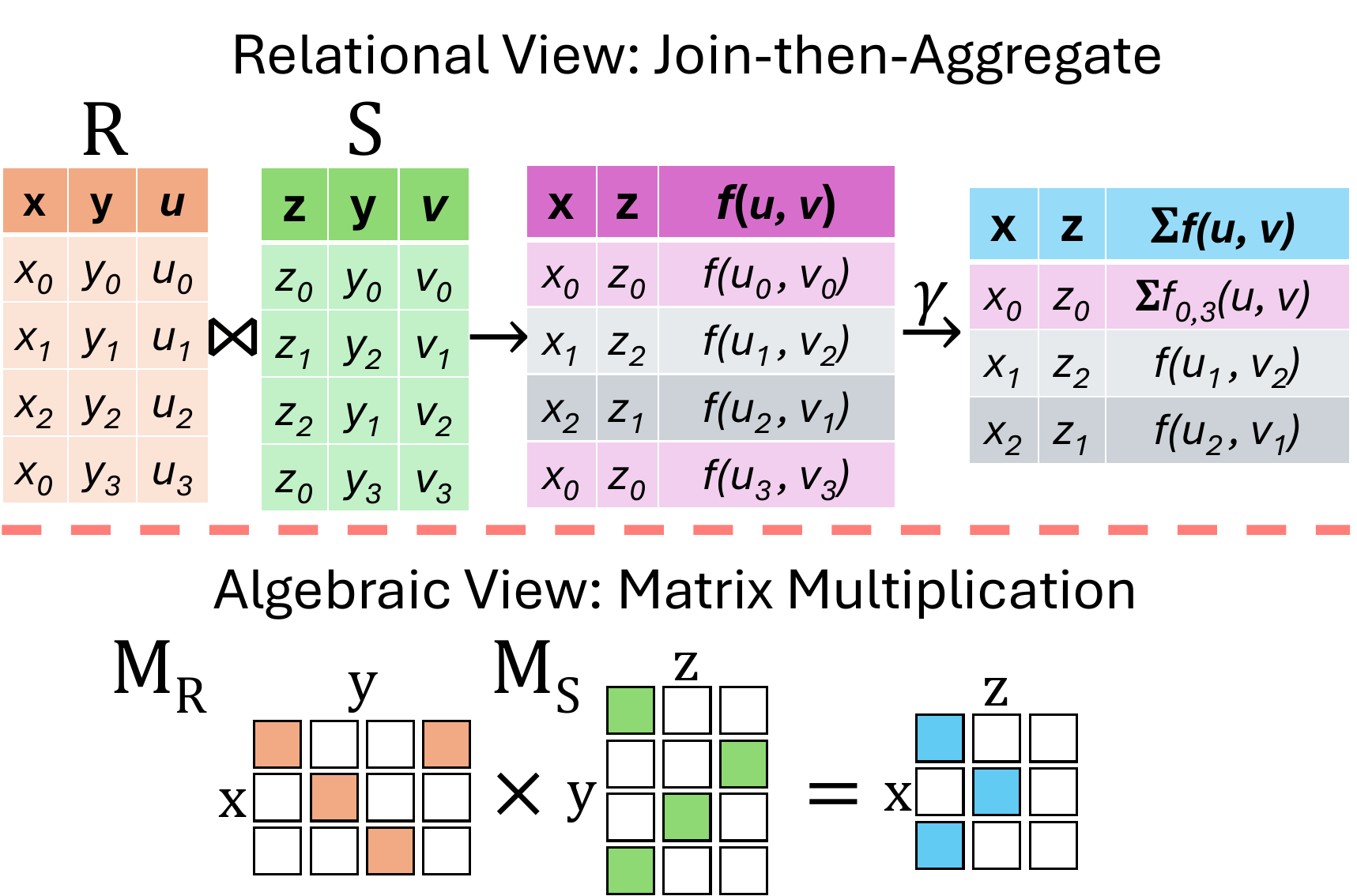}

	\caption{
		Comparing two approaches for evaluating a join-aggregate query, $``$ $SELECT$ $x$, $z$, $SUM(f(u, v))$ $FROM$ $R$, $S$ $WHERE$ $R.y = S.y$ $GROUP$ $BY$ $x$, $z$ $"$, where $f$ denotes a mathematical function.
		The ``relational view'' (top) joins tables $R(x, y, u)$ and $S(z, y, v)$ on attribute $y$ followed by an aggregation, a process that produces excessive intermediate results, such as $(x_0, z_0, f(u_0, v_0))$ and $(x_0, z_0, f(u_3, v_3))$, which are ultimately aggregated into one final group-by result (shaded in light pink).
		The ``algebraic view'' (bottom) models the operation as a matrix product $M_R \times M_S$ where $y$ serves as the shared dimension to compute the aggregate result directly, bypassing the materialization of costly intermediate results. %
	}

	\Description{}
	\label{fig:joinagg-vs-mm}
\end{figure}

	Matrix multiplication is a fundamental computational kernel, and its sparse variant, sparse general matrix-matrix multiplication (SpGEMM), is especially important in practice.
	The reason is that sparse matrices are a natural model for real-world graph data, which makes SpGEMM a core building block for applications such as graph analytics, entity resolution, and machine learning~\cite{DBLP:journals/tkde/ZhangLGT22,DBLP:journals/csur/GaoJCHWLW23,DBLP:conf/icde/ParkKJK023,jang2023sage,DBLP:journals/pacmmod/Lin0F0LM24,DBLP:journals/pvldb/SunHYYW11,DBLP:conf/icde/LuoGGPJ17}.
	Recently, SpGEMM has emerged as a transformative method for processing relational join-aggregate and join-project queries~\cite{deep2020fast,huang2022density,hu2020parallel,hu2025output,DBLP:conf/icde/SunKH23,DBLP:journals/dpd/SunKH25,DBLP:conf/ssdbm/SunK023,DBLP:conf/sigmod/HuL022,DBLP:journals/pacmmod/KimR26}.
	By treating relational tables as algebraic sparse matrices, entire query pipelines can be executed as unified algebraic operations. This enables aggregation or projection to be performed \textit{on-the-fly} alongside the join, effectively bypassing the prohibitive costs of materializing excessive intermediate results 
	(see Figure~\ref{fig:joinagg-vs-mm} for a comparison)
	.
	In all such cases, the performance of the application is dictated by the efficiency of the underlying parallel SpGEMM execution.
	Reflecting its critical role, numerous studies have investigated SpGEMM optimization, reaching a broad consensus that it is a fundamentally bandwidth-intensive kernel where throughput is ultimately governed by available memory bandwidth~\cite{DBLP:conf/spaa/GuMEA20,DBLP:journals/pc/NagasakaMAB19}.

	Though it is bandwidth-bound, running SpGEMM on a high-bandwidth platform does not necessarily translate to higher throughput
		~\footnote{While SpGEMM above a certain density threshold can be compute-limited, the growing compute-bandwidth gap~\cite{DBLP:journals/pvldb/LernerA24,yale75,micron23,DBLP:journals/micro/GholamiYKHMK24} makes compute progressively less likely than bandwidth to be the primary bottleneck.}
	.
	This is primarily attributed to the \emph{load imbalance} caused by the widely observed skew patterns in real-world workloads~\cite{DBLP:journals/pvldb/KimSCKNBLSD09,DBLP:conf/icde/VitorovicE016,DBLP:journals/pacmmod/LeeLCCLPPKK24,DBLP:conf/sc/YesilH0T20,DBLP:conf/sigmod/BlanasLP11,DBLP:journals/pvldb/SchmidtLBN25,tpcds}.
	For instance, in a typical OLAP ClickHouse benchmark~\cite{DBLP:journals/pvldb/SchulzeSYDM24}, the top $5\%$ of users generate $24.4\%$ ($47.8$ million) of all clicks, while the median number of clicks per user is only $2$. Similarly, in a common PageRank workload~\cite{webbase1m,DBLP:conf/sigmod/HuL022}, the top $1.8\%$ of pages receive more incoming links than the remaining $98.2\%$ combined.
	From a matrix perspective, such skewed key frequencies and power-law graph degrees translate into highly irregular nonzero structures, resulting in significant load imbalance.
	When running SpGEMM on these workloads, a small subset of cores becomes overloaded while others remain underutilized~\cite{DBLP:journals/pc/NagasakaMAB19,DBLP:conf/spaa/GuMEA20,DBLP:journals/csur/GaoJCHWLW23}.
	This not only limits compute efficiency but also prevents full utilization of the available memory bandwidth, leaving system resources idle, and therefore, rendering high-bandwidth platforms less effective.

	Moreover, the challenge of addressing load imbalance in skewed workloads is particularly critical and is further exacerbated by the ``memory scaling wall,'' where memory technology advancements lag behind those of processing units~\cite{yale75,micron23}.
	This trend results in decreasing memory bandwidth and capacity per processing core~\cite{DBLP:conf/vldb/WeisgutRSFTR25,DBLP:conf/vldb/Huang0LCHT24}, which further compounds the load imbalance problem for two reasons: 

	First, decreasing bandwidth-per-core has an acutely negative impact on SpGEMM workloads with load imbalance, where a few straggling threads drag down performance.
	Since these threads largely rely on core-level bandwidth, their overall performance is inevitably reduced due to the insufficient bandwidth-per-core caused by the ``memory scaling wall.''

\begin{figure}
	\centering

	\includegraphics[width=\linewidth]{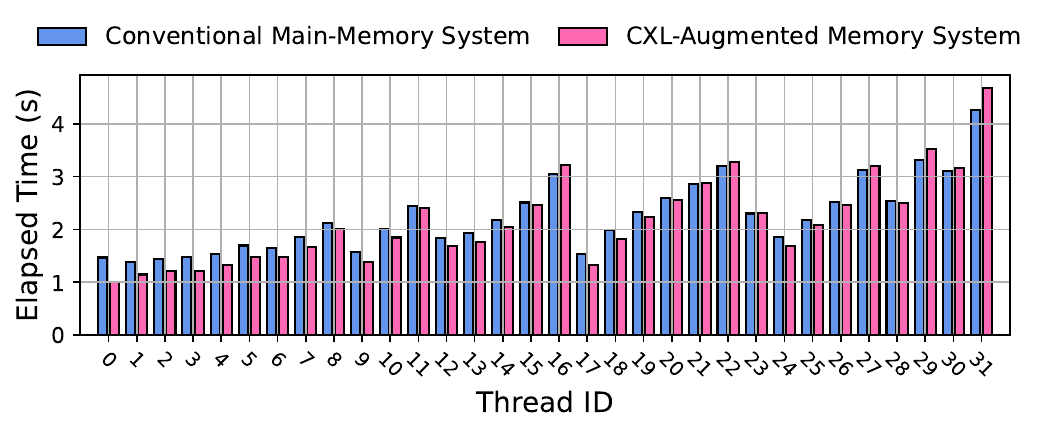}
	\vspace{-3em}
	\caption{
		Comparison of per-thread elapsed time for hashing-based SpGEMM~\cite{DBLP:journals/pc/NagasakaMAB19} on a G500 skewed matrix (scale=$17$, \textit{edgefactor}=$16$)~\cite{DBLP:conf/sdm/ChakrabartiZF04} between a conventional main-memory system ($28.50$~GB/s bandwidth) and a CXL-augmented system ($46.50$~GB/s bandwidth) (see \S~\ref{sec:exp} for experimental setups).
	}
	\Description{}
	\label{fig:numc-thr-g500-17-16}
\end{figure}

	Second, the prevalent approach to compensating for decreasing bandwidth-per-core, namely the adoption of heterogeneous memory expansion technologies like CXL~\cite{weisgut2025cxl,DBLP:conf/vldb/LebedevNUJA25,DBLP:conf/vldb/WeisgutRSFTR25,DBLP:conf/cidr/HuangLT26}, can inadvertently amplify this problem.
	Since CXL-attached memory typically exhibits higher access latency than local DRAM
		~\footnote{Since DRAM is the primary technology used in memory manufacturing, we use the terms ``DRAM'' and ``memory'' interchangeably throughout this paper.}
	, straggling threads that run disproportionately long are more exposed to CXL's latency penalty, further widening the gap between fast and slow threads.
	Figure~\ref{fig:numc-thr-g500-17-16} illustrates this using a hashing-based SpGEMM algorithm~\cite{DBLP:journals/pc/NagasakaMAB19} on a skewed workload, where per-thread runtimes vary due to load imbalance.~\footnote{We use the following notation for R-MAT graphs~\cite{DBLP:conf/sdm/ChakrabartiZF04}: \textbf{ER}, \textbf{SSCA}, and \textbf{G500} denote non-skewed Erd\H{o}s--R\'enyi, skewed SSCA, and highly skewed Graph500 graphs, respectively; \textit{scale} sets the number of rows to $2^{\textit{scale}}$, and \textit{edgefactor} sets the average nonzeros per row.}
	Because total runtime is governed by the slowest thread, the disproportionate latency on stragglers negates the aggregate bandwidth gain, yielding worse overall performance than pure DRAM. 
	This counterintuitive result is also reported in existing CXL studies~\cite{DBLP:conf/micro/SunYYKSHJALJ0AX23,DBLP:conf/eurosys/TangZZHYXLSHZCZ24,DBLP:conf/asplos/LiuHWBNJNL25,DBLP:conf/asplos/LiuXBAL26}.
	As a result, addressing load imbalance is becoming more imperative than ever.

	However, addressing load imbalance in SpGEMM is not trivial.
	While numerous strategies have been proposed~\cite{DBLP:conf/supercomputer/PatwarySSPAV0PP15,DBLP:journals/csur/GaoJCHWLW23,DBLP:conf/spaa/GuMEA20}, we maintain that existing approaches fall short due to two co-existing root causes: \emph{multi-dimensional skew} and \emph{cache contention}, both of which not only hinder load balancing but also degrade overall performance.

	First, existing methods fail to accommodate the \emph{multi-dimensional skew} inherent in sparse matrices.
	Current methods rely on static schemes that balance along a single metric (e.g., flop count) to minimize scheduling overhead.
	However, skew manifests across multiple dimensions, including computation cost, materialization overhead, and data locality, making it impossible for any single static metric to achieve balanced execution across all threads.

	Second, in skewed workloads, existing approaches suffer from significant cache contention, induced by dense substructures that are co-scheduled together.
	In SpGEMM, irregular sparsity forces practitioners to cache row-wise intermediates to reduce random access costs. 
	However, dense substructures within skewed matrices incur excessive irregular random entry access, which creates heavy cache pressure and even induces significant cache contention, which has long been overlooked as a key performance factor.
	This cache contention is particularly pronounced under simultaneous multithreading (SMT), which aims to increase computation parallelism but worsens the interference.

	In this paper, we address these challenges through a novel approach, \emph{\myappfull{} (\myapp{})}. 
	In contrast to de facto methods, \myapp{} revisits \emph{dynamic scheduling}, a strategy conducive to load balancing that is often dismissed due to concerns regarding scheduling overhead and poor performance~\cite{DBLP:conf/spaa/GuMEA20,DBLP:journals/pc/NagasakaMAB19}. 
	Through analysis (\S~\ref{subsec:dynamic}), we find that task granularity is the primary performance factor: a granularity that is too fine fails to preserve spatial locality and therefore suffers from excessive cache thrashing.
	Hence, we employ a morsel-wise dynamic scheduling strategy, where a ``morsel''\footnote{We adopt the term from Leis et al.~\cite{DBLP:conf/sigmod/LeisBK014}, where it denotes a fixed-size fragment of input data, whereas in our work it denotes a group of consecutive rows with a roughly fixed output size.} is defined as a basic scheduling unit of several consecutive matrix rows.
	\myapp{} ensures a morsel is large enough to maintain spatial locality, thereby greatly mitigating cache thrashing, and more importantly achieves load balance on skewed workloads through dynamic scheduling, effectively preventing computational idling and improving memory bandwidth utilization.

	To alleviate cache contention, \myapp{} introduces two lightweight and generalizable techniques: \emph{\myschfull{} (\mysch{})}, which mitigates cache interference by adaptively scheduling dense morsels, and \emph{\mydahafull{} (\mydaha{})}, which selectively enables SMT based on workload contention.
	These optimizations are simple in design, yet effective in practice, and easy to generalize to other sparse matrix kernels (e.g., SpMM and SpMV).
	Importantly, their performance benefits remain effective across heterogeneous bandwidth-boosted systems (e.g., CXL systems), ensuring that latency heterogeneity does not cause performance degradation.

	Furthermore, the morsel-wise scheduling strategy enables a \emph{\myaccumfull{} (\myaccum{})} for SpGEMM, where different accumulation algorithms can be selectively applied to morsels based on their intrinsic characteristics, thereby achieving more efficient execution and enhanced overall SpGEMM performance.

	The contributions of this paper are summarized as follows:
	\begin{enumerate}[wide, labelindent=0pt]
		\item We examine and characterize two co-existing root causes of performance degradation in skewed SpGEMM workloads: \emph{multi-dimensional skew}, where no single static metric can balance all threads, and \emph{cache contention}, where dense substructures induce severe on-chip interference that existing methods overlook.

		\item We develop a novel morsel-wise dynamic scheduling framework, \myappfull{} (\myapp{}), that systematically addresses both \emph{multi-dimensional skew} and \emph{cache contention} in SpGEMM.
		\myapp{} decouples skew dimensions through fixed-result-size morsels, integrates two key techniques, \myschfull{} (\mysch{}) and \mydahafull{} (\mydaha{}), to alleviate on-chip cache interference, and enables \myaccumfull{} (\myaccum{}) for morsel-wise accumulation adaptively.
		Our approach is straightforward in design, yet novel and generalizable, and effective in practice.

		\item We evaluate \myapp{} across a wide range of workloads, including matrix squaring, relational join-aggregate and join-project queries, graph analytics, and other sparse kernels, on both a main-memory and a CXL-augmented system, and show that it performs on par with state-of-the-art methods on non-skewed workloads while significantly outperforming them on skewed ones, addressing both computational idling and bandwidth underuse.
	\end{enumerate}

	The rest of the paper is organized as follows.
	We first present the background and motivation in \S~\ref{sec:bgmt}, followed by a description of \myapp{} in \S~\ref{sec:meth}.
	We then show the experimental results in \S~\ref{sec:exp} and discuss related work in \S~\ref{sec:rw}.
	Finally, we conclude the paper in \S~\ref{sec:concl}.

\section{Background and Motivation}
\label{sec:bgmt}
	We first review state-of-the-art SpGEMM algorithms, then analyze how multi-dimensional skew and cache contention degrade performance under skewed workloads, and finally revisit dynamic scheduling as a promising solution, motivating our \myapp{} approach.

	\subsection{The SpGEMM Kernel}
		\label{subsec:spgemm}

		SpGEMM computes the product $C = AB$, where $A$, $B$, and $C$ are all sparse matrices
		~\footnote{Throughout this paper, we refer to $A$ and $B$ in $C = AB$ as the left-hand side and right-hand side matrices, respectively.}
		.
		This form has proven effective across diverse large-scale workloads, including scientific computing, graph analytics, and more recently relational query processing~\cite{DBLP:conf/icppw/NagasakaMAB18,DBLP:journals/pc/NagasakaMAB19,hu2025output, DBLP:conf/icde/SunKH23}.
		To reduce memory footprint at scale, sparse matrix formats are widely used, with compressed sparse row (CSR) being the most common representation in existing works~\cite{intel-mkl,DBLP:journals/ijhpca/BulucG11,DBLP:conf/para/GilbertRS06,saad2003iterative,barrett1994templates}.
		However, due to the irregular and input-dependent sparsity patterns of sparse matrices, predicting the number and locations of nonzeros in the output matrix is often intricate, making memory pre-allocation challenging.
		Combined with irregular memory access patterns, these challenges make SpGEMM a fundamentally memory-intensive workload, with performance critically dependent on sustained memory bandwidth across cores.

		To address these challenges, SpGEMM is typically divided into two phases: a symbolic phase, which determines the nonzero structure of the output, and a numeric phase, which performs the actual computation based on that structure. 
		In addition, the irregular nature of sparsity makes it difficult to efficiently gather and accumulate scattered contributions to each output entry. 
		To tackle this, prior work has introduced a variety of accumulator designs to aggregate partial results during computation, including sorted variants such as heaps~\cite{DBLP:conf/icppw/NagasakaMAB18,DBLP:journals/pc/NagasakaMAB19}, SPA~\cite{DBLP:journals/toms/Gustavson78,gilbert1992sparse}, and expand-sort-compress (ESC)~\cite{dalton2015optimizing,bell2012exposing}, as well as non-sorted variants based on hash tables~\cite{DBLP:conf/icppw/NagasakaMAB18,DBLP:journals/pc/NagasakaMAB19,DBLP:conf/icpp/HongWMLXL24}.

		\newcommand{\Input}{\State \textbf{Input: }}
\newcommand{\Output}{\State \textbf{Output: }}
\algdef{SE}[PARFOR]{ParallelFor}{EndParallelFor}[1]%
	{\textbf{parallel for} #1\ \algorithmicdo}%
	{\algorithmicend\ \textbf{parallel for}}

\begin{algorithm}[t]
\captionsetup{skip=2pt}
\caption{Parallel Gustavson's Algorithm for SpGEMM}
\label{algo:gustavson}
	\begin{algorithmic}[1]
	\Input{CSR matrices $A \in \mathbb{R}^{m \times k}$ and $B \in \mathbb{R}^{k \times n}$}
	\Output{CSR matrix $C \in \mathbb{R}^{m \times n}$}
	\ParallelFor{$i \gets 0$ to $m - 1$} \Comment{Parallel over rows of $A$}
		\State Initialize empty \texttt{accumulator} for row $i$
		\ForAll{$(k, A_{ik}) \in$ row $i$ of $A$}
			\ForAll{$(j, B_{kj}) \in$ row $k$ of $B$}
				\If{$j \in$ \texttt{accumulator}}
					\State \texttt{accumulator[$j$]} $\mathrel{+}= A_{ik} \cdot B_{kj}$
				\Else
					\State \texttt{accumulator[$j$]} $\gets A_{ik} \cdot B_{kj}$
				\EndIf
			\EndFor
		\EndFor
		\State Store \texttt{accumulator} as row $i$ of $C$
	\EndParallelFor
	\end{algorithmic}
\end{algorithm}

		Building on these data representations, SpGEMM algorithms are commonly implemented using one of three formulations: inner-product~\cite{9065523}, outer-product~\cite{DBLP:conf/spaa/GuMEA20,DBLP:journals/siamsc/AkbudakA14}, or row-row paradigm (Gustavson's algorithm)~\cite{DBLP:journals/toms/Gustavson78}. 
		Among these, Gustavson's algorithm operates in a row-wise manner, allowing each row of the left-hand side matrix ($A$) to be computed independently, enabling straightforward thread-level parallelization without the need for inter-thread synchronization (see Algorithm~\ref{algo:gustavson}). 
		It is this high degree of parallelism, combined with its efficient memory access patterns, that has made Gustavson's algorithm a dominant formulation for SpGEMM across a wide range of computing platforms.

		Following Gustavson's algorithm, several load-balancing techniques have been proposed~\cite{DBLP:conf/icpp/BulucG08,DBLP:journals/topc/BallardDKS16,DBLP:journals/siamsc/AkbudakA14,DBLP:conf/sc/TrotterELTDIU23,DBLP:journals/tpds/ChenC22,DBLP:conf/ipps/AraiSYOI16}, with static scheduling based on a flop-count scheme emerging as the dominant strategy~\cite{DBLP:journals/pc/NagasakaMAB19,DBLP:conf/icppw/NagasakaMAB18}.
		This method partitions the rows of matrix~$A$ into contiguous regions, each assigned approximately the same number of floating-point operations (i.e., additions plus multiplications).
		To achieve this, a lightweight preprocessing pass scans matrices~$A$ and~$B$ to estimate the flop count for each output row of~$C$ without performing actual calculations.
		Based on these estimates, rows are grouped into partitions such that each thread receives a workload with a nearly equal number of floating-point operations.
		This approach enables static, flop-count-balanced scheduling across threads while avoiding runtime synchronization.

	\subsection{The Flop-Count-Based Scheme Under Skew}
	\label{subsec:skew}
		Despite its simplicity and widespread adoption~\cite{DBLP:journals/ijhpca/BulucG11,DBLP:journals/pc/NagasakaMAB19}, the flop-count-based scheme is fundamentally limited and fails to achieve effective workload balance under skewed patterns.
		We identify two root causes, an inherent ignoring of multi-dimensional skew and the performance penalty of overlooking cache contention, both of which hurt load balancing and degrade bandwidth utilization.

		\textbf{(1)} \textbf{Ignoring multi-dimensional skew.}
		Skewed SpGEMM workloads can exhibit imbalance across multiple dimensions, such as submatrix density and row-wise nonzero ($nnz$) distribution. 
		However, the flop-count-based scheme equalizes only the total number of floating-point operations, ignoring other forms of skew. 
		As a result, one thread may receive a few dense rows (with high $nnz$), while another is assigned many sparse rows. 
		Although these assignments may have similar total flop counts, their structural differences lead to divergent runtime behaviors. 
		Dense rows, in particular, induce ``clustering''
		~\footnote{We use the term ``clustering'' to describe this effect, following the notion of ``primary clustering'' in open-addressing hashing.}
		effects during processing, which manifest as temporal bursts of memory and cache pressure. 
		These effects significantly degrade performance, and the extent of impact varies by accumulation strategy~\cite{DBLP:conf/icppw/NagasakaMAB18,dalton2015optimizing,gilbert1992sparse}.
		Therefore, the scheme is inherently unable to account for these skew dimensions.

		\begin{figure}
	\centering

	\includegraphics[width=\linewidth
	] %
	{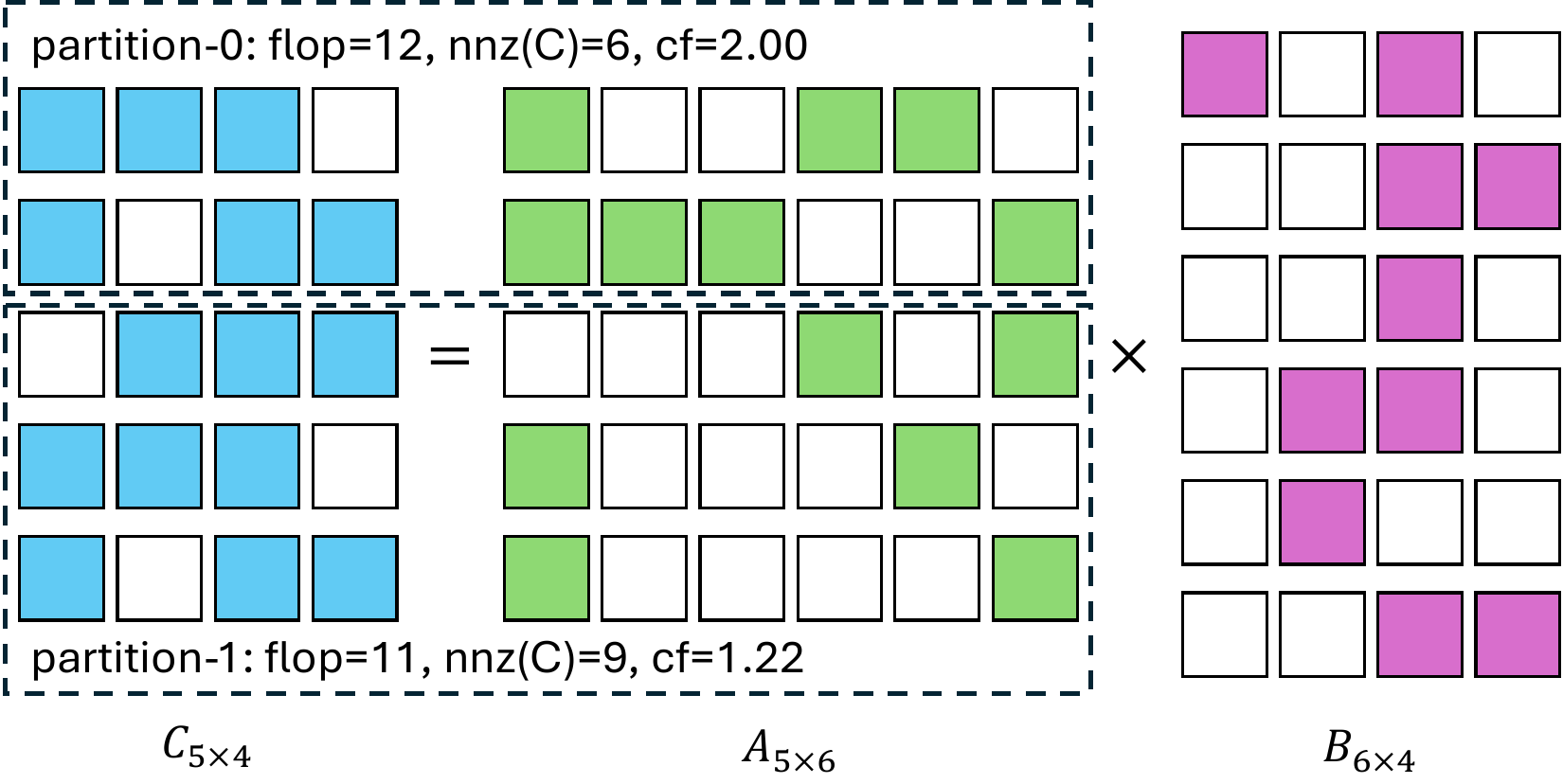}
	\vspace{-2.5em}
	\caption{An illustration of $C = AB$ in SpGEMM (colored cells denote nonzero elements), where the two dashed boxes represent partitions with comparable flop counts but differing numbers of output nonzeros and compression factors.}

	\Description{}
	\label{fig:cf}
\end{figure}

\begin{figure}
	\centering

	\includegraphics[width=\linewidth]{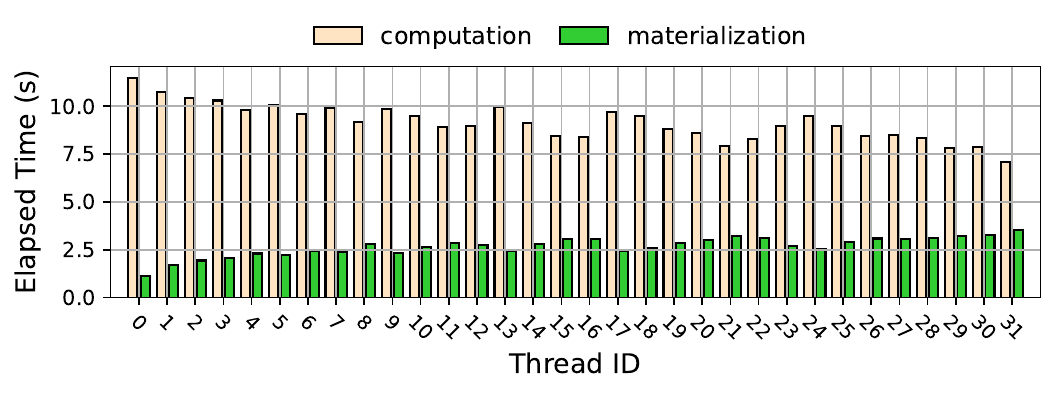}
	\vspace{-3em}
	\caption{
		Thread-level computation time vs. materialization time for hashing-based SpGEMM on a G500 skewed matrix (scale=$19$, \textit{edgefactor}=$8$).
	}
	\Description{}
	\label{fig:hash-numc-comp-write}
\end{figure}

\begin{figure*}
	\centering

	\includegraphics[width=\linewidth]{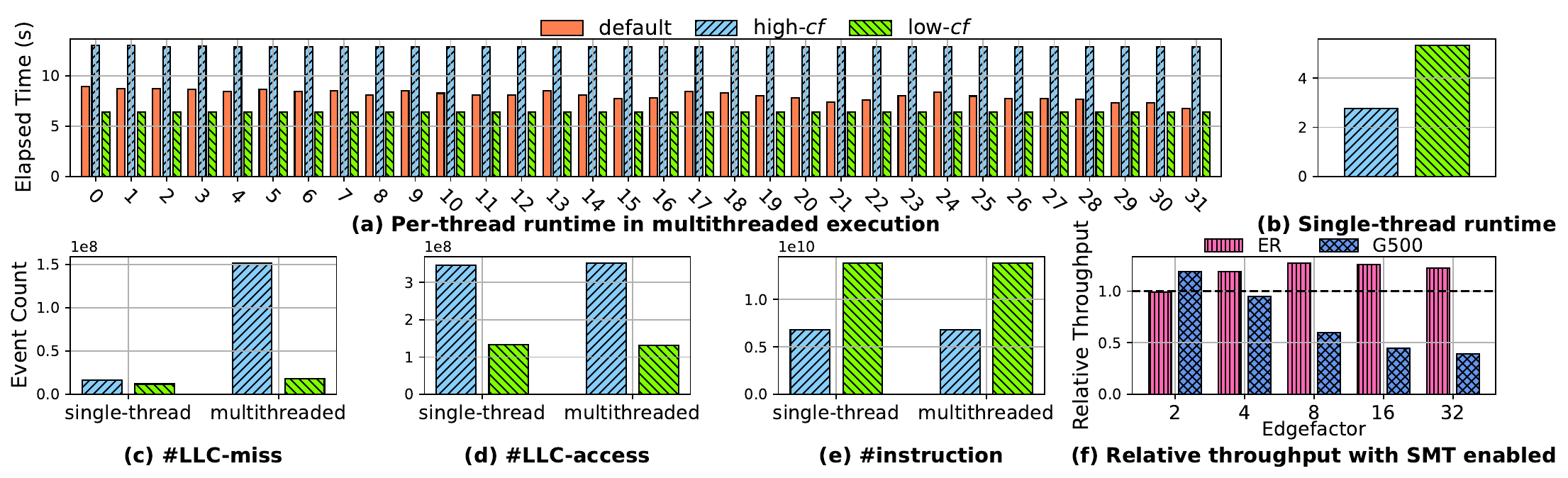}
	\vspace{-2.5em}
	\caption{
		(a) Per-thread computation time of HASH-based SpGEMM on a G500 skewed matrix (scale=$19$, \textit{edgefactor}=$8$) under three settings: all threads execute their default partitions; all execute a high-$cf$ partition (thread-$0$'s); all execute a low-$cf$ partition (thread-$31$'s).
		(b) Computation time of the low-$cf$ and high-$cf$ partitions under single-threaded execution.  
		(c)--(e) LLC misses, LLC accesses, and instruction counts for the low-$cf$ and high-$cf$ partitions under both single-threaded and multithreaded execution.  
		(f) Normalized throughput of HASH-based SpGEMM with SMT enabled on ER and G500 matrices (scale=$18$) across varying \textit{edgefactor} values. The black dashed line marks the relative throughput without SMT.
	}

	\Description{}
	\label{fig:cont}
\end{figure*}

		Additionally, thread-level partitions often exhibit different compression factors ($cf$)
		~\footnote{The compression factor ($cf$) is defined as the number of flop counts divided by the $nnz$ of the result matrix $C$, i.e., $cf = \frac{flop}{nnz(C)}$ ~\cite{DBLP:conf/spaa/GuMEA20}.}
		, which leads to divergent trends in computation and materialization times, making it challenging to effectively balance these two workload components.
		Consider Figure~\ref{fig:cf}: using Gustavson's algorithm, partition-$0$ (the upper two rows of $A$) requires $12$ flop counts
		~\footnote{Only multiplications are counted, as the number of additions is identical.}
		, producing $6$ nonzeros; partition-$1$ (the bottom three rows) requires $11$ flop counts but produces $9$ nonzeros. 
		Despite both involving comparable amounts of computation, the second has a larger output, inducing more materialization cost.

		Empirical data on a skewed workload confirms this disparity, as Figure~\ref{fig:hash-numc-comp-write} shows: materialization time varies across threads, with computation time also exhibiting thread-level variation, but in the opposite trend. %
		Moreover, these complexities are further compounded by hardware variability: performance sensitivity to the relative costs of computation versus memory access differs considerably across hardware platforms. 
		This architectural variance makes it exceedingly difficult for purely flop-count-based schemes to establish a load balance that is both universally effective and readily generalizable across diverse machines.
		Thus, such schemes are inherently limited in delivering robust load balancing.

		\textbf{(2)} \textbf{Overlooking cache contention.}
		Existing methods also overlook the performance impact of cache contention during parallel execution.
		In Gustavson's algorithm, each thread maintains an exclusive row-level accumulator whose size scales with the nonzero density of its assigned rows.
		Reading Figure~\ref{fig:cf} row-wise, each row of $A$ gathers the $B$ rows indexed by its nonzeros into that accumulator, so $cf$ measures the nonzeros of $B$ streamed per accumulator entry; a high-$cf$ row thus places significantly greater pressure on shared caches than a low-$cf$ row.
		As high-$cf$ partitions generally contain more high-$cf$ rows and Gustavson's algorithm is typically parallelized row-wise, they are more prone to inter-thread cache interference than low-$cf$ partitions.
		Figure~\ref{fig:cont}(a) shows this effect using the same skewed matrix as in Figure~\ref{fig:hash-numc-comp-write}. 
		When all threads compute their assigned default partitions, the runtime of a high-$cf$ partition (originally assigned to thread-$0$ in the flop-count-based scheme) increases substantially: from $2.76$ seconds in single-threaded execution (see Figure~\ref{fig:cont}(b)) to $8.93$ seconds under default multithreaded partitioning, and further to $13.01$ seconds if all threads execute this high-$cf$ partition.
		Note that no sharing benefit arises in this case, as each thread operates on its own exclusive accumulator; instead, these accumulators compete for the same shared cache, amplifying interference as more threads process high-$cf$ regions.
		In contrast, a low-$cf$ partition (originally assigned to thread-$31$) sees only a modest runtime increase, from $5.36$ to $6.73$ seconds.
		
		Further profiling reveals that cache contention is the root cause of this disparity: Figure~\ref{fig:cont}(c)--(e) shows a sharp rise in last-level cache (LLC) misses for the high-$cf$ region, while instruction count and LLC accesses remain relatively stable, confirming cache contention as the primary bottleneck
		~\footnote{Write-out time is excluded to directly address computation-level contention.}
		. 

		The problem is further exacerbated under simultaneous multithreading (SMT): Figure~\ref{fig:cont}(f) shows that non-skewed workloads (e.g., ER matrices with typically low $cf$) benefit from SMT, whereas skewed workloads (e.g., G500 matrices with high-$cf$ subregions) experience increasing throughput degradation as density grows, since higher density amplifies the skew and thus intensifies SMT-induced contention.
		This underscores the importance of contention-aware scheduling; neglecting this factor not only compromises load balance but also degrades overall SpGEMM efficiency.

\begin{figure}
	\centering

	\includegraphics[width=0.9\linewidth]{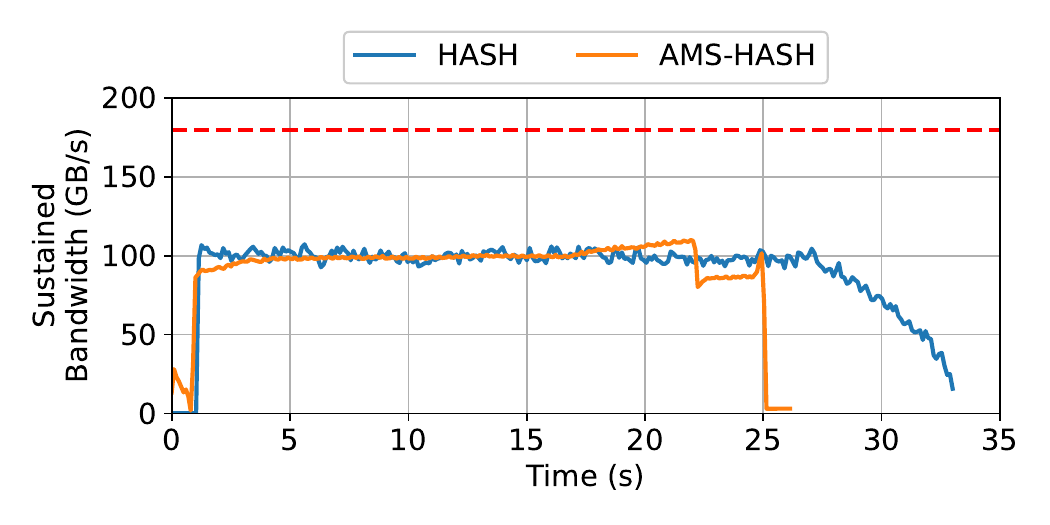}
	\vspace{-2em}
	\caption{
		Comparison of sustained bandwidth of the numeric phase between flop-count-based hashing-based SpGEMM and the \myapp{} variant on a G500 skewed workload (scale=$19$, \textit{edgefactor}=$16$); the horizontal red dashed line denotes system memory bandwidth ($179.50$ GB/s).
	}
	\Description{}
	\label{fig:susbw-tai}
\end{figure}

		Ignoring multi-dimensional skew and overlooking cache contention together compound into a macro-level penalty on system-wide bandwidth utilization.
		As Figure~\ref{fig:susbw-tai} illustrates on a skewed workload, bandwidth resources are only fully utilized during the initial burst of parallel execution; as straggler cores emerge, system-wide bandwidth consumption starts to drop, leaving expensive memory resources idle.
		This renders prevailing bandwidth expansion technologies, such as CXL, far less effective, as the slowest thread governs total runtime, and its prolonged latency~\cite{DBLP:conf/micro/SunYYKSHJALJ0AX23} further widens the gap despite the increased aggregate bandwidth available.
		In contrast, a contention-aware load-balancing approach (our \myapp{} method) sustains markedly higher bandwidth execution.

	\subsection{Revisiting Dynamic Scheduling}
		\label{subsec:dynamic}

\begin{figure}
	\centering
	\begin{subfigure}[b]{0.47\linewidth}
			\centering
			\includegraphics[width=\linewidth]{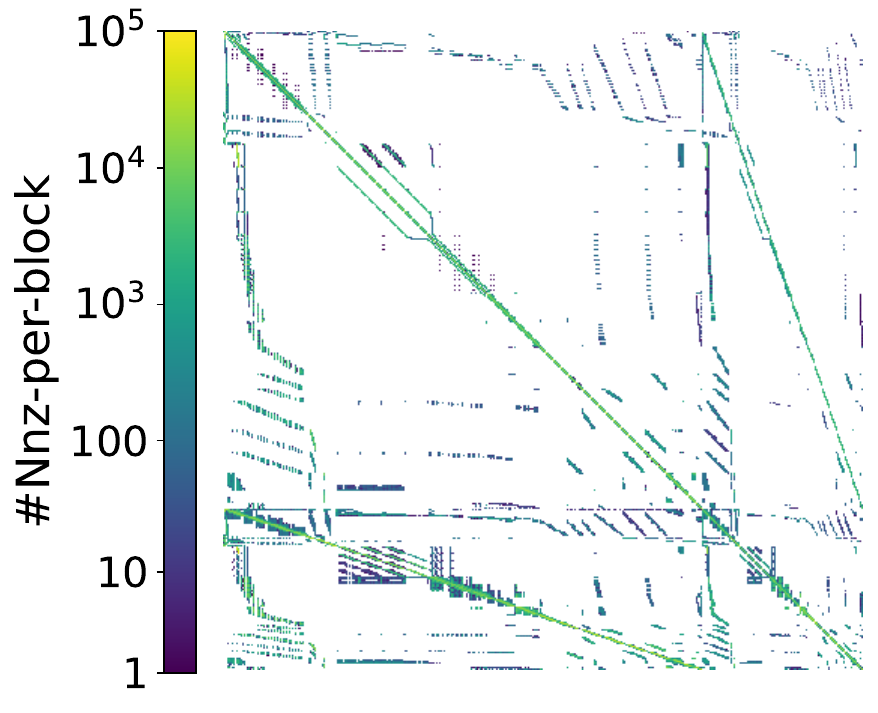}
			\vspace{-1.75em}
			\caption{ {\footnotesize NNZ-pattern  } }
			\label{fig:vas_stokes_1m-nnz}
	\end{subfigure}
	\begin{subfigure}[b]{0.49\linewidth}
			\centering
			\includegraphics[width=\linewidth]{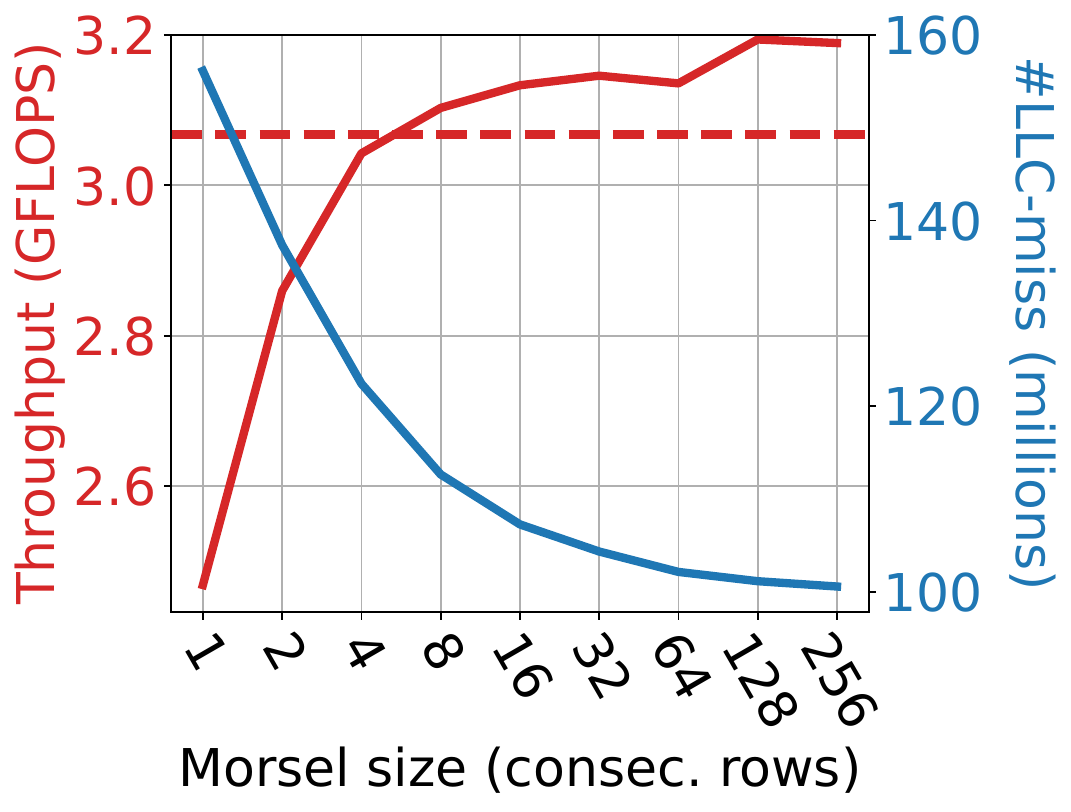}
			\vspace{-1.75em}
			\caption{ \footnotesize Throughput \& LLC-miss}
			\label{fig:vas_stokes_1m-tps-llc-miss}
	\end{subfigure}
	\vspace{-1em}
	\caption{
		(a) Nonzero pattern of the \texttt{vas\_stokes\_1M} matrix.
		(b) Throughput and LLC misses of hashing-based SpGEMM on \texttt{vas\_stokes\_1M} using dynamic scheduling with varying morsel sizes (measured in consecutive rows per morsel); the horizontal red dashed line indicates the throughput of the flop-count-based scheme.
	}
	\Description{}
	\label{fig:vas_stokes_1m}
\end{figure}

		Having established that both multi-dimensional skew and cache contention undermine existing scheduling strategies, we now revisit the scheduling paradigm that enables our solution.
		Specifically, we revisit \emph{dynamic scheduling}~\cite{DBLP:journals/pc/NagasakaMAB19,DBLP:conf/supercomputer/PatwarySSPAV0PP15}: rows of $A$ are placed into a shared task pool, and threads repeatedly fetch and process one row at a time, achieving nearly $100\%$ load balance. 
		While simple and theoretically well-balanced, this approach has been shown to underperform compared to flop-count-based scheduling and has largely been dismissed in recent literature, which attributes the performance penalty primarily to the synchronization overhead associated with dynamic task dispatch~\cite{DBLP:journals/pc/NagasakaMAB19,DBLP:conf/spaa/GuMEA20}.
		However, we contend that it is task granularity, rather than synchronization overhead, that plays the dominant role in determining performance. 
		When task size is carefully managed, dynamic scheduling can match, if not outperform, flop-count-based schemes.

		Take a real-world SpGEMM workload, \texttt{vas\_stokes\_1M}~\cite{vasstokes1m,DBLP:journals/toms/DavisH11}, for example.
		The matrix exhibits clear spatial locality across rows, particularly within blocks of consecutive rows (Figure~\ref{fig:vas_stokes_1m}(a)).
		We benchmark the performance of hashing-based SpGEMM using dynamic scheduling with varying task granularities (i.e., the number of consecutive rows per task).
		Figure~\ref{fig:vas_stokes_1m}(b) shows that throughput improves with increasing granularity, plateauing around $128$ rows, and surpasses its flop-count-based counterpart due to better load balance.
		Further analysis of LLC miss counts reveals that larger task sizes result in fewer misses, suggesting better preservation of spatial locality.
		It is worth noting that many real-world matrices exhibit strong spatial locality in their nonzero patterns, and prior works have introduced several offline reordering algorithms to enhance locality for downstream operations~\cite{DBLP:journals/tpds/ChenC22,DBLP:conf/ipps/AraiSYOI16,DBLP:conf/sc/TrotterELTDIU23}. 
		Based on these observations, we maintain that dynamic scheduling remains a practical and efficient strategy as long as spatial locality is effectively preserved. 
		We therefore build on this insight for the design of our \myapp{} approach.

\section{Methodology}
\label{sec:meth}

	This section presents our proposed framework, \myappfull{} (\myapp{}). 
	We begin with the core concept of morsel-wise dynamic scheduling, our strategy for addressing multi-dimensional skew and achieving efficient load balancing in SpGEMM. 
	Building upon this foundation, we present two key techniques, \myschfull{} (\mysch{}) and \mydahafull{} (\mydaha{}), designed specifically to alleviate on-chip cache contention. 
	Furthermore, we explain how the morsel-wise framework enables our \myaccumfull{} (\myaccum{}) approach for optimizing SpGEMM execution. 
	Finally, we discuss three broader implications of \myapp{}, namely the interplay between load balance and cache efficiency that governs the benefit of bandwidth-expanded memory, the carry-over of the underlying design principle to relational and graph processing, and the extension of \myapp{} to other sparse matrix multiplication kernels.

	We note that \myapp{} targets the numeric phase of SpGEMM, which dominates both computation and memory traffic.
	While the same ideas can be applied to the symbolic phase, the benefits are less pronounced; for simplicity, we use standard dynamic scheduling with fixed-size morsels for that phase.

	Before detailing our techniques, we briefly digress to outline three key design principles that an efficient SpGEMM solution should satisfy, all of which are met by our proposed \myapp{}.
	\begin{enumerate} [wide, labelindent=0pt]
		\item \textbf{Simple and lightweight.}  
		SpGEMM has been extensively optimized over decades~\cite{DBLP:journals/csur/GaoJCHWLW23}, and any added complexity can degrade its performance.
		This is especially critical when SpGEMM is repeatedly invoked as a core kernel in real-world systems~\cite{DBLP:journals/ijhpca/BulucG11,intel-mkl,eigen,scipy2020,DBLP:journals/pc/NagasakaMAB19,DBLP:journals/pacmmod/Lin0F0LM24}.
		Thus, any proposed method must be simple and lightweight to avoid introducing significant execution penalties.

		\item \textbf{Non-disruptive to non-skewed workloads.}  
		While skewed patterns are common in real-world workloads, non-skewed distributions (e.g., uniform distribution, symmetric distribution) also arise frequently. 
		A design that benefits skewed cases but penalizes non-skewed ones forces practitioners to make case-dependent trade-offs, increasing system complexity and operational overhead.

		\item \textbf{Generalizable across algorithms and platforms.}
		An effective SpGEMM solution should generalize across algorithmic variants and hardware platforms.
		In practice, SpGEMM is used in diverse contexts with varying requirements, e.g., numerical computing often demands sorted output~\cite{dalton2015optimizing}, while relational and graph queries typically accept non-sorted results~\cite{DBLP:conf/para/GilbertRS06,DBLP:conf/ipps/AraiSYOI16,DBLP:journals/tpds/ChenC22}.
		Also, it should adapt to different hardware architectures, where sensitivities to compute and memory bottlenecks vary.
		A practical solution must therefore accommodate both, supporting sorted and non-sorted pipelines while maintaining robust performance across platforms.
	\end{enumerate}

	\subsection{Morsel-Wise Dynamic Scheduling}
	\label{subsec:morsel}

		We previously demonstrated that, by using morsels composed of a sufficient number of consecutive rows, dynamic scheduling is able to effectively preserve spatial locality in sparse matrices.
		This method significantly reduces LLC misses with negligible synchronization overhead (\S~\ref{subsec:dynamic}). 
		We therefore adopt this line of design for SpGEMM execution.

		While scheduling work in morsels of a fixed row count helps reduce cache thrashing, we argue that this approach leaves room for further improvement, as it does not address the \emph{multi-dimensional skew} identified in \S~\ref{subsec:skew}. Specifically, it has two limitations:
		(1) Fixed-row-count morsels do not account for differences in flop count or result $nnz$, making it difficult to balance computation and materialization costs. 
		This limitation is particularly pronounced in skewed workloads, where high-$cf$ submatrices incur high computational expense while others involve substantial write-out overhead. 
		Due to such disparities in two dimensions, modeling inter-thread cache contention becomes challenging. 
		(2) This fixed-row strategy also fails to control the size of the resulting output. 
		A single morsel containing a few dense rows, particularly in skewed workloads, can produce more data than the CPU cache can accommodate, making it challenging to construct a cache-aware analysis model~\cite{DBLP:conf/sigmod/SchuhCD16,DBLP:conf/sigmod/0002SLLF15,DBLP:conf/icpp/HongWMLXL24}.

		To address these issues, we propose scheduling morsels based on a fixed result size, i.e., a target $nnz(C)$ per morsel. 
		Recall that SpGEMM consists of two phases, where the symbolic phase determines the per-row nonzero structure. 
		We thus leverage this information to form morsels of consecutive rows in an adaptive manner, such that each morsel contains approximately the same number of output nonzeros. 
		By fixing the materialization cost per morsel, this approach normalizes one dimension of skew, allowing us to focus on the remaining computational cost dimension.
		It also decouples $cf$ from the output size that flop-count-based partitioning leaves uncontrolled: with $nnz(C)$ fixed, $cf$ varies solely with the flop count.
		Additionally, to facilitate on-chip cache modeling, we limit the morsel size to half of the L2 cache capacity~\footnote{
		We empirically find that morsel sizes ranging from $\frac{1}{8}$ to $\frac{1}{2}$ of the L2 cache capacity offer the best trade-off between performance and cache contention.
		}.

		This strategy is simple and efficient, and it forms the foundation for our subsequent designs.
		More importantly, it introduces no additional algorithmic phase, fully aligning with our design principle of being \textbf{simple and lightweight}.

	\subsection{ \myschfullcap{} }
	\label{subsec:calb}
		We now elaborate on our alleviation of inter-thread on-chip cache contention, which can degrade SpGEMM performance even under balanced workload distribution. 
		Since we form morsels based on a fixed result size, we can easily compute the compression factor ($cf$) for each morsel. 
		Recall from \S~\ref{subsec:skew} that contention originates at the row level: a high-$cf$ row streams more nonzeros of $B$ per accumulator entry (see Figure~\ref{fig:cf}), and is thus contention-prone regardless of how rows are grouped.
		With $nnz(C)$ fixed per morsel, a high-$cf$ morsel likewise carries more computation against an identical footprint, and remains more prone to contention than a low-$cf$ one.
		Therefore, we aim to co-execute high-$cf$ and low-$cf$ morsels concurrently, alleviating cache contention and reducing susceptibility to inter-thread interference.

		\begin{figure}
	\centering

	\includegraphics[width=\linewidth
	]{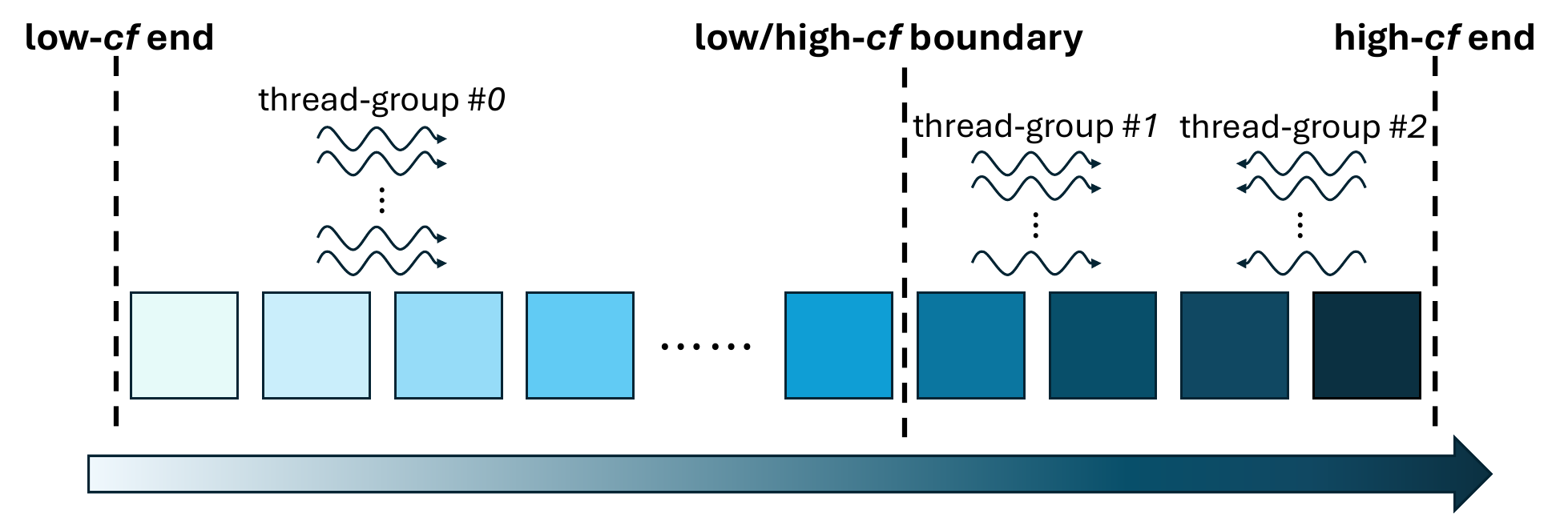}
	\vspace{-2.5em}
	\caption{
		Thread scheduling with \myschfull{} (\mysch{}). Darker colors indicate higher compression factors ($cf$).
	}
	\Description{}
	\label{fig:calb}
\end{figure} 

		To that end, we first sort all morsels in ascending order of $cf$ and empirically set a boundary between low-$cf$ and high-$cf$ regions, typically between the $80$th and $95$th percentile.
		This is motivated by the observation that most skewed workloads follow a power-law distribution~\cite{DBLP:journals/tpds/ChenC22,DBLP:conf/sc/TrotterELTDIU23,DBLP:conf/ipps/AraiSYOI16,DBLP:conf/sc/YesilH0T20}, where a small fraction of dense rows contribute to the majority of the workload. 
		We then divide the available threads into three groups: 
		thread group $\#0$ starts from the low-$cf$ end and progresses toward the high-$cf$ end; 
		group $\#1$ begins at the low/high-$cf$ boundary and proceeds toward the high-$cf$ end; 
		group $\#2$ starts at the high-$cf$ end and moves backward toward the boundary (see Figure~\ref{fig:calb}). 
		The majority of threads are allocated to group $\#0$ to handle low-$cf$ tasks, while smaller subsets are assigned to groups $\#1$ and $\#2$ for high-$cf$ regions. 
		This arrangement ensures that only a limited number of high-$cf$ morsels are processed concurrently. 
		Dividing the remaining threads into groups $\#1$ and $\#2$ further helps mitigate cache contention by avoiding simultaneous execution of multiple contention-prone morsels at the high-$cf$ end. 
		Meanwhile, co-running low-$cf$ morsels, which are less sensitive to cache interference, alongside high-$cf$ morsels helps absorb contention effects and improves overall execution efficiency.

		While this scheme alleviates cache contention, it does not guarantee load balance. 
		Some thread groups may finish early and become idle, leading to underutilization of computational and memory bandwidth resources. 
		To address this, we allow early-finished threads to assist other active groups. 
		Specifically, if group $\#0$ reaches the boundary early, it merges with group $\#1$ to continue processing toward the high-$cf$ end.
		Likewise, early-finished threads from groups $\#1$ and $\#2$ join group $\#0$ to help complete tasks in the low-$cf$ region. 
		This cooperative mechanism keeps all threads active, ensuring efficient use of computation and memory bandwidth.

		The \mysch{} scheme is conceptually simple, yet novel and highly effective. 
		To the best of our knowledge, it is the first SpGEMM load-balancing technique that explicitly considers on-chip 
		cache contention as a scheduling factor and thereby delivers tangible performance benefits.
		Moreover, the method is platform-agnostic and can be seamlessly integrated into any SpGEMM algorithm based on Gustavson's formulation, adhering to the design principle of being \textbf{generalizable across algorithms and platforms}.

	\subsection{\mydahafullcap{}}
	\label{subsec:smtact}
		While SpGEMM is generally considered bandwidth-bound, its computational phase still incurs a non-negligible overhead. 
		This overhead can typically be mitigated by leveraging simultaneous multithreading (SMT), which increases the logical core count to effectively improve computational throughput.
		However, as shown in Figure~\ref{fig:cont}(f), this performance benefit from SMT is primarily observed in non-skewed workloads; in contrast, skewed workloads degrade significantly when SMT is enabled.

		We attribute this disparity to the increased on-chip cache pressure from SMT. 
		Although SMT increases the logical thread count, it does not expand the physical cache. 
		Consequently, high-$cf$ morsels suffer from intensified cache contention, an issue that is particularly exacerbated in skewed workloads where $cf$ values are disproportionately large for a few morsels. 
		These specific morsels are highly vulnerable to SMT-induced interference, which in turn leads to substantial performance degradation.

		To mitigate this issue while preserving the computational benefits of SMT for non-skewed workloads, we adopt a selective activation strategy of SMT, namely \mydahafull{} (\mydaha{}).
		Specifically, we set a threshold based on the highest $cf$ observed.
		SMT is disabled if the workload contains any morsels exceeding this threshold, and enabled otherwise.
		This strategy ensures that SMT is applied only when the workload is less prone to cache contention
		~\footnote{We also experiment with using quantiles of the $cf$ distribution instead of the maximum, and observe comparable performance with an appropriately chosen threshold.}.
		The approach is simple yet effective, as it balances the computational benefits of SMT against the risk of elevated cache contention, aligning with our second design principle of being \textbf{non-disruptive to non-skewed workloads}.

	\subsection{\myaccumfullcap{}}
	\label{subsec:hyb}
		Due to the inherently sparse and irregular nature of SpGEMM, computing a single output row generates multiple intermediate sparse products.
		These products must then be merged, a process that requires appropriate accumulators to mitigate two challenges: severe irregular memory access patterns and the high cost of managing collisions for identical column indices.
		Recognizing these challenges, numerous works have focused on developing efficient accumulators~\cite{DBLP:journals/pc/NagasakaMAB19,dalton2015optimizing,gilbert1992sparse}. 
		However, the effectiveness of any given accumulation strategy varies dramatically based on the characteristics of the underlying matrix morsel. 
		This makes a static, one-size-fits-all accumulator inherently suboptimal.

		To address this, we leverage the morsel-wise framework of \myapp{} to introduce a \myaccumfull{} (\myaccum{}) that dynamically selects the most suitable accumulator for each morsel.
		Specifically, we make this choice based on the aforementioned morsel-wise $cf$ value, selectively choosing between two widely used accumulators: hash tables (HASH)~\cite{DBLP:journals/pc/NagasakaMAB19,DBLP:conf/icppw/NagasakaMAB18}, which use linear probing and multiplicative hashing to reduce collisions, and expand-sort-compress (ESC)~\cite{dalton2015optimizing,DBLP:conf/spaa/GuMEA20}, which relies on highly optimized radix sorting to alleviate common logarithmic sorting overheads.

\begin{figure}
	\centering
	\includegraphics[width=0.66\linewidth]{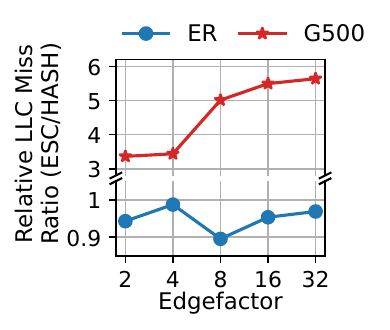}
	\caption{
		The LLC miss ratio of ESC normalized to HASH (i.e., LLC Misses of ESC / LLC Misses of HASH), evaluated on ER and G500 datasets at Scale $18$ with varying edgefactors (higher edgefactors suggest denser workloads).
	}
	\Description{}
	\label{fig:llc-miss-ratio}
\end{figure}

		The choice between these two accumulators is highly dependent on the morsel-wise $cf$. 
		As ESC generates sorted output, it executes more instructions than HASH. 
		However, this instruction overhead is not necessarily detrimental, particularly for low-$cf$ morsels.
		For such morsels, the number of intermediate products is similar to the final number of nonzeros, leading to a moderate cost for merging and sorting. 
		In contrast, HASH becomes more populated for low-$cf$ morsels, and the overhead of hash computation and probing becomes more pronounced. 
		For high-$cf$ morsels, however, HASH is the better choice, as a massive number of intermediates are hashed into a few slots, greatly reducing potential cache misses. 
		Conversely, ESC must pay a significant cache miss cost to merge a large volume of intermediate results into a small final output. 
		Microbenchmark results in Figure~\ref{fig:llc-miss-ratio} corroborate this finding: the ratio of ESC's LLC misses to HASH's is substantially higher on skewed G500 workloads with high $cf$ than on non-skewed ER workloads.

		Therefore, we empirically identify a $cf$ threshold, using ESC for morsels with a $cf$ below this value and opting for HASH for morsels above it~\footnote{We find that a $cf$ of $1.5$--$1.7$ provides a robust threshold for selecting the appropriate accumulator.}. 
		This morsel-wise selection also offers an architectural benefit: selecting the accumulator on a per-morsel basis alleviates the instruction-level branch-misprediction costs that finer-grained (e.g., per-row) selection would incur.

		It is worth noting that by including HASH, our \myaccum{} approach may produce a non-sorted result matrix,
		which is acceptable for applications like relational queries and graph analytics that do not require sorted output~\cite{huang2022density,DBLP:conf/icppw/NagasakaMAB18}.
		However, if sorted output is necessary, \myaccum{} can be easily adapted:
		instead of choosing between HASH and ESC, it can be configured to choose the efficient strategy from a set of sorted accumulators, such as ESC, HEAP~\cite{DBLP:journals/pc/NagasakaMAB19}, SPA~\cite{gilbert1992sparse}, or even a hash-then-sort approach~\cite{DBLP:conf/icpp/ChengLL023}, to ensure a fully sorted output.
		Consequently, this adaptive approach ensures each morsel is processed with the most suitable accumulator, leading to better overall SpGEMM performance regardless of the output ordering requirement.

	\subsection{Implications and Broader Applicability}
	\label{subsec:ext}
		Having presented \myapp{}'s components, we now step back to consider what they imply beyond the SpGEMM kernel itself.
			\myapp{} highlights an important interplay between load balance and cache efficiency.
		Static partitioning preserves locality but leaves cores idle under skew, thereby underutilizing memory bandwidth; dynamic scheduling keeps cores busy but raises cache contention once the working sets of co-executing morsels exceed the shared cache.
		Effective scheduling must therefore maintain enough concurrency to saturate the memory system while limiting avoidable cache traffic, which is what \myapp{} achieves by scheduling morsels from cheap metadata.
		This balance is particularly important for bandwidth-expanded memory~\cite{DBLP:conf/vldb/WeisgutRSFTR25}, where utilizing the additional bandwidth requires enough cores to issue concurrent memory requests. By keeping cores busy while reducing cache interference, \myapp{} sustains memory bandwidth beyond the local-DRAM peak (\S~\ref{subsec:cxl}).

		The same design principle can apply beyond SpGEMM when lightweight metadata is available to estimate each morsel's processing cost and demand on the shared cache.
		For SpGEMM, the symbolic phase provides this information directly.
		Other operators can obtain approximate metadata through a preliminary statistics pass~\cite{DBLP:conf/icde/BalkesenTAO13,DBLP:conf/sigmod/ChenY17} or sampling~\cite{DBLP:journals/toms/Vitter85,DBLP:conf/icde/WangC20}; exact estimates are unnecessary because \mysch{} uses only the relative ordering of morsels by cache demand and \mydaha{} only tests whether a threshold is crossed.
		This creates opportunities to apply contention-aware morsel scheduling to other operators over relational and graph data with skewed, cache-resident state~\cite{amossen2009faster,deep2020fast,huang2022density,DBLP:conf/para/GilbertRS06,DBLP:journals/pvldb/SunHYYW11}.
		For example, a morsel-driven engine~\cite{DBLP:conf/sigmod/LeisBK014,DBLP:journals/pvldb/DursunBCSG19} could use partition-size estimates to approximate cache demand when scheduling hash-aggregation or hash-join tasks over skewed keys.

		A direct and notable extension of \myapp{}, beyond SpGEMM and the relational query and graph operators it underpins, is to apply it atop other sparse matrix multiplication kernels, e.g., sparse matrix-dense matrix multiplication (SpMM) and sparse matrix-vector multiplication (SpMV), which multiply a sparse matrix by a dense operand.
		Their output is dense and of fixed size, which renders \myaccum{} unnecessary and makes morsel formation by a target result size meaningless; we instead group consecutive rows by a target $nnz$ of the left-hand operand.
		Although the skew thereby reduces to a single dimension (flop count), load imbalance and cache contention persist, as morsels containing many short rows and those containing a few long rows differ markedly in data reuse~\cite{DBLP:journals/csur/XiaoYZLCL24,DBLP:journals/corr/abs-2404-06047}. Therefore, \mysch{} and \mydaha{} can address these challenges in SpMM and SpMV using the same principles as in SpGEMM~\cite{DBLP:conf/icde/LiKZGT25}.
		Since most SpMM and SpMV solutions likewise follow a Gustavson-style row-wise formulation, \myapp{} can be adapted to these kernels in a straightforward manner.

\section{Experimental Evaluation}
\label{sec:exp}
	This section evaluates \myapp{} in addressing multi-dimensional skew and on-chip cache contention.
	We begin with the experimental setup (\S~\ref{subsec:setup}), then benchmark \myapp{} on matrix squaring across diverse workloads (\S~\ref{subsec:perf}) and dissect \mysch{} and \mydaha{} to validate their contention alleviation (\S~\ref{subsec:dissect}).
	We further evaluate \myapp{} on a bandwidth-expanded CXL system to show improved bandwidth utilization (\S~\ref{subsec:cxl}).
	We then demonstrate its broad applicability on other sparse matrix multiplication kernels (\S~\ref{subsec:perf-spmmv}), on relational join-aggregate and join-project queries (\S~\ref{subsec:perf-joinagg}), and on graph-analytic workloads (\S~\ref{subsec:perf-graph}), including PathSim meta-path construction~\cite{DBLP:journals/pvldb/SunHYYW11}, breadth-first search, and triangle counting.
	We close with a discussion (\S~\ref{subsec:disc}) that distills the key findings across these workloads and platforms, and relates them back to the design rationale of \myapp{}.

	\subsection{Experimental Setup}
	\label{subsec:setup}

\begin{table}[t]
\centering
\caption{Summary of 20 representative matrices evaluated.}
\label{tab:tamu}
	\resizebox{\columnwidth}{!}{%
	\begin{tabular}{lrrrrrr}
		\toprule
		\textbf{Matrix} & \textbf{row($A$)} & \textbf{nnz($A$)} & \textbf{gini($A$)$^\dagger$} & \textbf{flop$^\ddagger$} & \textbf{nnz($A^2$)} & \textbf{cf} \\
		\midrule

		hugetric-00020 	& 7.13M	& 21.37M	& 0.0003	& 64.08M	& 44.61M	& 1.4363	\\
		GL7d18 	& 1.96M	& 35.60M	& 0.0395	& 890.29M	& 843.47M	& 1.0555	\\
		333SP 	& 3.72M	& 22.22M	& 0.0559	& 134.79M	& 71.97M	& 1.8730	\\
		NLR 	& 4.17M	& 24.98M	& 0.0736	& 152.77M	& 82.02M	& 1.8626	\\
		M6 	& 3.51M	& 21.01M	& 0.0742	& 128.49M	& 68.99M	& 1.8625	\\
		AS365 	& 3.80M	& 22.74M	& 0.0765	& 138.91M	& 74.62M	& 1.8616	\\
		auto 	& 448.70K	& 6.63M	& 0.1016	& 101.25M	& 33.06M	& 3.0634	\\
		delaunay\_n23 	& 8.39M	& 50.34M	& 0.1220	& 316.97M	& 173.67M	& 1.8251	\\
		nv2 	& 1.46M	& 37.48M	& 0.1401	& 1.28B	& 292.14M	& 4.3475	\\
		Freescale1 	& 3.43M	& 17.06M	& 0.1858	& 90.33M	& 38.31M	& 2.3580	\\
		gsm\_106857 	& 589.45K	& 21.76M	& 0.2240	& 947.26M	& 132.40M	& 7.1546	\\
		amazon-2008 	& 735.33K	& 10.32M	& 0.2856	& 218.17M	& 63.16M	& 3.4546	\\
		webbase-1M 	& 1.01M	& 3.11M	& 0.3455	& 1.37B	& 1.31B	& 1.0475	\\
		vas\_stokes\_2M 	& 2.15M	& 65.13M	& 0.3771	& 6.95B	& 1.17B	& 5.9452	\\
		rel9 	& 9.89M	& 23.67M	& 0.4016	& 2.10B	& 2.04B	& 1.0300	\\
		vas\_stokes\_1M 	& 1.10M	& 34.77M	& 0.4081	& 4.39B	& 721.63M	& 6.0776	\\
		patents 	& 3.78M	& 14.98M	& 0.6585	& 178.30M	& 132.12M	& 1.3496	\\
		cit-Patents 	& 3.78M	& 16.52M	& 0.6844	& 239.46M	& 162.35M	& 1.4749	\\
		NotreDame\_www 	& 325.73K	& 929.85K	& 0.7825	& 418.01M	& 288.98M	& 1.4465	\\
		web-NotreDame 	& 325.73K	& 1.50M	& 0.8523	& 503.57M	& 293.20M	& 1.7175	\\		

		\bottomrule
		\multicolumn{7}{l}{
			\parbox[t]{\linewidth}{%
				$^\dagger$ Gini coefficient~\cite{DBLP:conf/websci/KunegisP12} of the row-wise nonzeros in $A$.\\
				$^\ddagger$ Only multiplications are counted, as additions are equal in number.
			}%
		}
	\end{tabular}
	}
\end{table}

		We conduct experiments on a single-socket machine running Linux kernel 6.11.0. 
		This system is equipped with an Intel Xeon® Gold 6430 CPU, which features $32$ physical cores and supports $64$ logical threads via simultaneous multithreading (SMT)
		~\footnote{We disable Intel Speed Select Technology - Base Frequency to ensure equal performance for all physical cores~\cite{LinuxKernelSST}.}
		.  
		Each core includes a $48$~KB L1 data cache, a $32$~KB L1 instruction cache, and a $2$~MB L2 cache, while all cores share a $60$~MB last-level cache (LLC). 
		The system comprises $192$~GB of DDR5 DRAM, providing $179.50$~GB/s of memory bandwidth.
		We also conduct experiments on a CXL-augmented memory system to validate the generalizability of \myapp{}, with the setup details deferred to \S~\ref{subsec:cxl}.

		Following prior settings~\cite{DBLP:conf/icppw/NagasakaMAB18,DBLP:conf/icpp/BulucG08,DBLP:journals/ijhpca/BulucG11,DBLP:journals/siamsc/AzadBBDGSTW16,DBLP:journals/siamsc/BulucG12}, we conduct our primary performance analysis using R-MAT matrix squaring workloads~\cite{DBLP:conf/sdm/ChakrabartiZF04,wsvr,DBLP:conf/icde/GaoLSLH25}.
		We employ three representative sets of R-MAT matrices to model different workload characteristics: ER ($a=b=c=d=0.25$), SSCA ($a=0.60, b=c=0.13, d=0.14$), and G500 ($a=0.57, b=c=0.19, d=0.05$), which represent non-skewed, skewed, and highly skewed workloads, respectively.
		All R-MAT generators support control over size and density via two parameters: \textit{scale} (determining the number of rows, i.e., $2^{\textit{scale}}$) and \textit{edgefactor} (defining the average nonzeros per row).
		We sweep \textit{scale} over $19$--$21$ for ER, $18$--$20$ for SSCA, and $17$--$19$ for G500, with \textit{edgefactor} over $2$--$32$ in all three.
		For SSCA and G500 workloads, higher \textit{edgefactors} amplify multi-dimensional skew and cache contention (\S~\ref{subsec:skew}).
		Additionally, we evaluate SpGEMM performance using real-world workloads from the SuiteSparse Matrix Benchmark~\cite{DBLP:journals/toms/DavisH11} to validate the robustness of \myapp{}.
		We perform matrix squaring on $119$ SuiteSparse matrices, with row counts ranging from $100$K to $10$M and nonzero counts from $146$K to $316$M.
		We choose this range so that even the smallest matrix exceeds the $2$~MB private L2 cache capacity%
		~\footnote{Storing a matrix of $n$ rows and $\mathit{nnz}$ nonzeros in CSR with $8$-byte indices and double-precision values takes $8(n+1)+16\,\mathit{nnz}$ bytes, corresponding to approximately $3$~MB at the lower end of this range and $5.1$~GB at the upper end.}.
		This ensures that all workloads exercise the cache hierarchy beyond private L2, while keeping the largest matrices small enough to complete on our platform without exhausting memory.
		For a clear presentation, we present results for $20$ representative matrices (Table~\ref{tab:tamu}), ordered by their Gini coefficient%
		~\footnote{The Gini coefficient measures inequality in the row-wise nonzero distribution; a value of $0$ indicates uniform distribution, while values approaching $1$ indicate extreme skew. Coefficients above $0.30$ generally indicate skewed workloads~\cite{DBLP:conf/websci/KunegisP12,DBLP:journals/toms/DavisH11}.}%
		, as other matrices show similar trends.
		Beyond matrix squaring, we evaluate \myapp{} on SpMM and SpMV (\S~\ref{subsec:perf-spmmv}), relational join-aggregate and join-project queries (\S~\ref{subsec:perf-joinagg}), and graph-analytic workloads, including PathSim meta-path construction, multi-source breadth-first search, and triangle counting (\S~\ref{subsec:perf-graph}). The corresponding workloads and experimental setups are detailed in these subsections.

		We integrate our \myapp{} framework into two prevalent Gustavson-based SpGEMM algorithms: the non-sorted HASH algorithm~\cite{DBLP:conf/icppw/NagasakaMAB18,DBLP:journals/pc/NagasakaMAB19} and the sorted ESC algorithm~\cite{dalton2015optimizing,bell2012exposing}.
		These are referred to as \myapp{}-HASH and \myapp{}-ESC, respectively
		~\footnote{Sorted algorithms generally incur more overhead to produce row-wise sorted output.}
		.
		Moreover, we implement our \myaccumfull{} (\myaccum{}) as \myapp{}-\myaccum{}, which adaptively selects between HASH and ESC accumulators.
		We empirically tune the parameters for \myapp{}-based approaches, configuring morsel size at half the L2 cache size, and distributing threads using a $60\%$-$25\%$-$15\%$ ratio across groups $\#0$, $\#1$, and $\#2$, respectively, for optimal \mysch{} performance.

		\definecolor{groupbg}{RGB}{237,234,226}      %
\definecolor{propbg}{RGB}{248,237,232}        %
\definecolor{ablbg}{RGB}{253,245,240}         %
\definecolor{accentred}{RGB}{139,32,32}       %
\definecolor{badgebase}{RGB}{230,227,218}     %
\definecolor{badgecomp}{RGB}{220,228,236}     %
\definecolor{badgeprop}{RGB}{245,228,224}     %

\begin{table}[t]
	\centering
	\caption[Summary of all evaluated SpGEMM methods]{Summary of all evaluated SpGEMM methods}
	\label{tab:spgemms}
	\renewcommand{\arraystretch}{1.2}
	\setlength{\tabcolsep}{2pt} 
	\footnotesize
	
	\begin{tabularx}{\columnwidth}{@{} >{\ttfamily\raggedright\arraybackslash}p{1.05cm} >{\raggedright\arraybackslash}p{1.05cm} >{\centering\arraybackslash}p{0.75cm} X @{}} 
		\toprule
		\normalfont\textbf{Method} & \textbf{Category} & \textbf{Sorted} & \textbf{Description} \\
		\midrule
		\rowcolor{groupbg}
		\multicolumn{4}{l}{\textit{Algorithmic baselines (built upon by our AMS methods)}} \\
		HASH~\cite{DBLP:journals/pc/NagasakaMAB19} & Baseline & \xmark & Hash table accumulator with flop-count-based scheduler. \\
		ESC~\cite{dalton2015optimizing}  & Baseline & \cmark & Radix-sort (expand-sort-compress) accumulator with flop-count-based scheduler. \\
		\midrule
		\rowcolor{groupbg}
		\multicolumn{4}{l}{\textit{External / industry competitors}} \\
		HEAP~\cite{DBLP:journals/pc/NagasakaMAB19} & Competitor & \cmark & Heap accumulator with flop-count-based scheduler. \\
		PB~\cite{DBLP:conf/spaa/GuMEA20}   & Competitor & \cmark & Outer-product SpGEMM flow with flop-count-based scheduler. \\
		MKL~\cite{intel-mkl}  & Competitor & \xmark & Intel oneMKL SpGEMM — non-sorted variant. \\
		MKLS~\cite{intel-mkl} & Competitor & \cmark & Intel oneMKL SpGEMM — sorted variant. \\
		\midrule
		\rowcolor{groupbg}
		\multicolumn{4}{l}{\textit{Proposed \myapp{}-based methods}} \\
		\rowcolor{ablbg}
		\myapp{}-HASH & Proposed & \xmark & \myapp{} approach using a HASH accumulator.  \\
		\rowcolor{ablbg}
		\myapp{}-ESC  & Proposed & \cmark & \myapp{} approach using an ESC accumulator. \\
		\rowcolor{propbg}
		\myapp{}-HYB  & Proposed & \xmark & \myapp{} with per-morsel adaptive HASH/ESC selection.  \\
		\bottomrule
	\end{tabularx}
\end{table}

		We compare our methods against several state-of-the-art baselines (see Table~\ref{tab:spgemms} for an overview).
		These include: 
		(1) the same HASH and ESC algorithms driven by a state-of-the-art flop-count-based scheduler; 
		(2) the sorted HEAP algorithm~\cite{DBLP:conf/icppw/NagasakaMAB18,DBLP:journals/pc/NagasakaMAB19}, which exploits the implicit column ordering of input matrices to reduce sorting overhead at the cost of increased memory usage; 
		(3) the outer-product-based PB method~\cite{DBLP:conf/spaa/GuMEA20}; 
		and (4) the de facto industry solution, Intel MKL~\cite{intel-mkl}, including both its sorted (MKLS) and non-sorted (MKL) variants.
		While other SpGEMM algorithms, such as SPA~\cite{gilbert1992sparse} and SaSpGEMM~\cite{DBLP:conf/icpp/HongWMLXL24}, were also evaluated, we omit their results for conciseness as they consistently underperformed.

		All implementations are compiled using \texttt{GCC-11.5.0} with the \texttt{-O3} optimization flag and parallelized with OpenMP~\cite{dagum1998openmp}.
		Baselines are manually tuned for optimal performance, which includes enabling SMT for non-skewed workloads and disabling it for skewed ones.
		Our \myapp{}-based methods are tuned with equal care, sweeping the thread group sizes, the low/high-$cf$ boundary, and the \mydaha{} threshold.
		As in prior works, throughput, measured in GFLOPS (giga floating-point operations per second), serves as the primary evaluation metric.
		Unless a measurement is explicitly labeled as numeric-phase profiling, all reported runtimes and throughputs cover the complete SpGEMM pipeline, i.e., the symbolic phase, prefix-sum computation and output allocation, \myapp{} preparation, the numeric phase, and output finalization.
		All reported results are averaged over $10$ runs~\footnote{The source code, data, and/or other artifacts have been made available at \url{\vldbavailabilityurl}.}.

	\subsection{Matrix Squaring Performance Evaluation}
	\label{subsec:perf}

\begin{figure*}
	\centering

	\includegraphics[width=\linewidth]{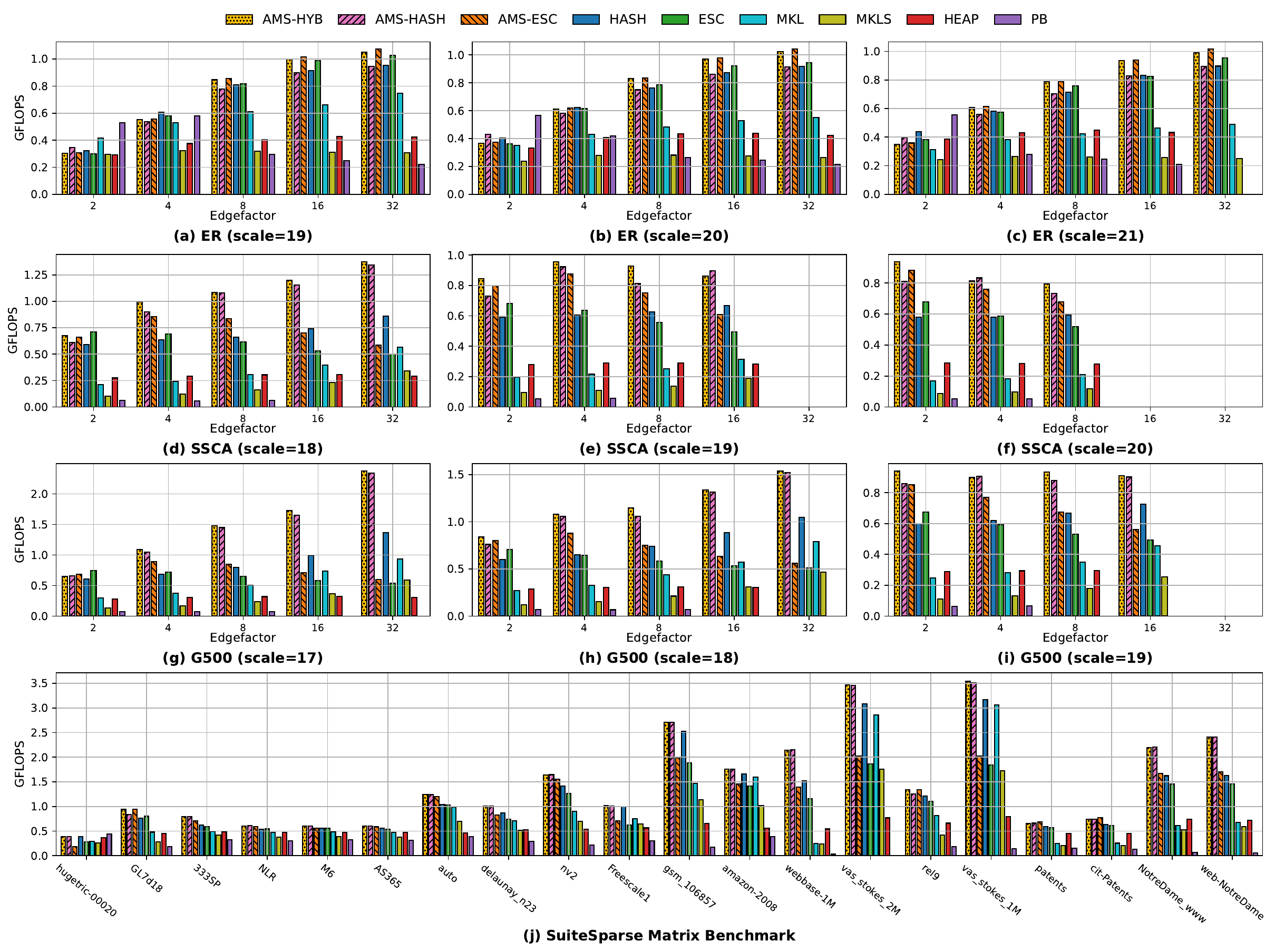}

	\caption{
		SpGEMM performance evaluation. 
		Subfigures (a)–(i) present results for R-MAT matrices (ER, SSCA, G500) across varying scales and \textit{edgefactor} values, representing changes in matrix dimension and density, respectively.
		Subfigure (j) displays results for 20 representative matrices from the SuiteSparse collection, ordered on the x-axis by their Gini coefficient (ascending), which correlates with increased workload skew. 
		Empty bars signify out-of-memory (OOM) failures.
	}
	\Description{}
	\label{fig:rmat-tamu}
\end{figure*}

		Figure~\ref{fig:rmat-tamu} presents the matrix squaring performance on R-MAT and SuiteSparse workloads, and shows that our \myapp{}-based algorithms consistently achieve high performance across this broad spectrum.
		Specifically, the proposed \myapp{}-\myaccum{} is generally the superior solution: it maintains performance comparable to \myapp{}-HASH in skewed workloads (SSCA and G500) while remaining as competitive as \myapp{}-ESC in non-skewed, high-\textit{edgefactor} workloads (ER).
		This outcome validates both the effectiveness of our \myaccum{} design and the soundness of our $cf$-based criterion for adaptively selecting the best morsel-wise accumulator for efficient multiplication.

		Beyond the superior performance of \myapp{}-\myaccum{}, the \myapp{} framework itself significantly boosts both its non-sorted (\myapp{}-HASH) and sorted (\myapp{}-ESC) variants.
		These methods perform comparably to their respective baselines on ER matrices but deliver a substantial performance advantage on the skewed SSCA and G500 matrices.
		For the non-sorted variant, the gains are larger on the more skewed G500 workloads than on SSCA at matched \textit{scale} and \textit{edgefactor}, where the two generators differ only in their R-MAT probabilities.
		For example, at \textit{scale}~$19$ and \textit{edgefactor}~$2$, \myapp{}-HASH achieves speedups of $1.44\times$ on G500 and $1.23\times$ on SSCA over the HASH baseline.
		The sorted \myapp{}-ESC variant also benefits on skewed workloads, although its gains are smaller than those of \myapp{}-HASH, as discussed below.
		Together, these results demonstrate that \myapp{}'s dynamic scheduling, when paired with appropriately sized morsels, effectively addresses multi-dimensional skew without sacrificing performance on non-skewed workloads.
		The strong results on ER matrices (Figure~\ref{fig:rmat-tamu}(a)–(c)) also validate the effectiveness of \mydaha{}, which selectively enables SMT in low-contention scenarios to enhance computational throughput, a critical factor in non-skewed workloads.

		Comparing the impact of \myapp{} across different algorithms, we observe that the performance improvement is more pronounced in HASH than in ESC. 
		This is because ESC relies on radix sort, which has a complexity of $O(\omega \cdot n)$ and correlates directly with the flop count. 
		Although HASH-based methods also exhibit $O(n)$ complexity in hash table lookups, their efficiency is more susceptible to ``clustering'' effects, making flop-count-based models less effective in accurately estimating cost and balancing workloads.
		In contrast, our \myapp{} approach is agnostic to the underlying SpGEMM algorithm.
		For both non-sorted and sorted algorithms, \myapp{} achieves performance gains of up to $81.01\%$ and $36.03\%$, respectively, demonstrating its general applicability across algorithmic variants.

\begin{figure*}
	\centering

	\includegraphics[width=\linewidth]{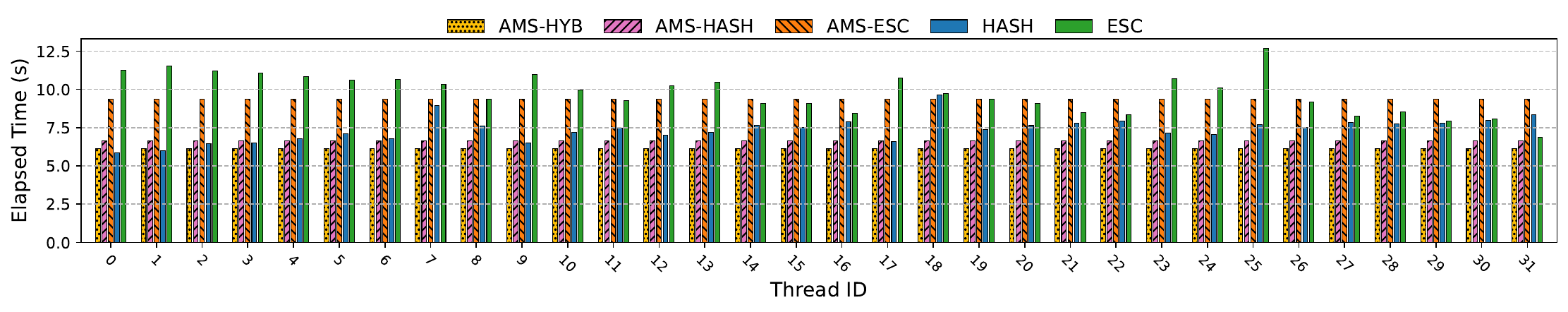}

	\caption{
		Per-thread elapsed time of the numeric phase comparison of \myapp{}-based variants (\myapp{}-\myaccum{}, \myapp{}-HASH, \myapp{}-ESC) and their baseline counterparts (HASH, ESC) on a G500 skewed matrix at \textit{scale}~$19$ and \textit{edgefactor}~$8$.
	}
	\Description{}
	\label{fig:g500_19_8_numc_thr}
\end{figure*}

		To further illustrate the source of these gains, Figure~\ref{fig:g500_19_8_numc_thr} drills down into the per-thread elapsed time, comparing \myapp{}-based methods against their respective baselines on a representative skewed workload.
		The baseline HASH and ESC exhibit significant thread-level runtime variance, confirming that flop-count-based scheduling fails to balance the actual workload under skew.
		In contrast, \myapp{}-based variants achieve substantially more uniform per-thread runtimes, demonstrating that our scheduling framework effectively addresses the load imbalance problem inherent in skewed workloads.

		Despite these advantages, Figure~\ref{fig:rmat-tamu} shows that \myapp{}-based algorithms may occasionally underperform their flop-count-based counterparts in highly sparse cases (e.g., G500 with \textit{scale}~$17$ and \textit{edgefactor}~$2$), primarily due to overhead from $cf$-based morsel sorting. 
		Nevertheless, this penalty diminishes as matrix density increases, underscoring the lightweight nature of \myapp{} in addressing load imbalance.

		Turning to the competing methods, a consistent performance ranking emerges across the tested configurations.
		The baseline HASH and ESC algorithms trail their \myapp{}-enhanced counterparts, and the de facto industry solution, Intel MKL, falls behind those baselines in turn: \myapp{}-\myaccum{} surpasses the non-sorted MKL by up to $3.83\times$ and the sorted MKLS by up to $8.45\times$, while \myapp{}-ESC delivers up to $7.66\times$ over MKLS.
		HEAP and PB rank last, with more unfinished runs owing to their notably larger memory demands.
		The one exception is PB on highly sparse non-skewed workloads (ER matrices, \textit{edgefactor}~$2$), where its outer-product-based formulation maximizes data reuse~\cite{DBLP:conf/spaa/GuMEA20}, an advantage that quickly diminishes at higher \textit{edgefactor} values and thus implies limited applicability across diverse workloads.
		These observations underscore the need for adaptive SpGEMM processing, as no single formulation dominates across the density spectrum.

		The SuiteSparse results (Figure~\ref{fig:rmat-tamu}(j)) show the same performance ranking on real-world matrices as on the synthetic R-MAT matrices.
		\myapp{}-\myaccum{} outperforms every competitor on all $20$ matrices, except for one matrix against HASH and one against PB.
		Consistent with the R-MAT results, \myapp{}-HASH and \myapp{}-ESC outperform their flop-count-based counterparts on $19$ of the $20$ matrices, by up to $47.46\%$ and $25.92\%$, respectively.
		Compared with the other methods, \myapp{}-\myaccum{} achieves geometric-mean speedups of $1.16\times$ over HASH, $1.37\times$ over ESC, $1.81\times$ over MKL, $2.24\times$ over HEAP, and $2.41\times$ over MKLS.
		Its largest advantage is over PB, with a geometric-mean speedup of $6.08\times$; PB is also the only method that fails to finish on one of the $20$ matrices.
		These results on real-world matrices show that \myapp{}'s benefits extend beyond the synthetic R-MAT workloads.

	\subsection{Dissecting \myapp{}: \mysch{} and \mydaha{}}
	\label{subsec:dissect}

		\subsubsection{Effectiveness of \mysch{}}

\begin{figure}
	\centering

	\includegraphics[width=\linewidth]{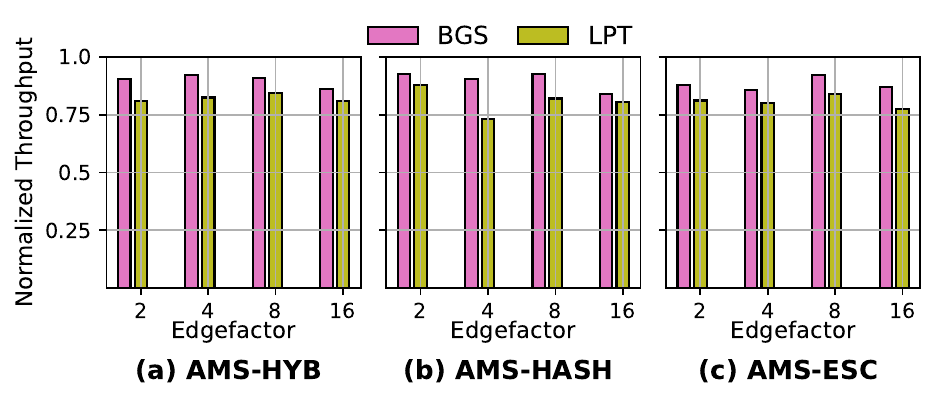}
	\caption{
		Throughput comparison of BGS and LPT scheduling strategies against \mysch{}. 
		The y-axis shows throughput on G500 (scale=$19$) workloads, normalized to \mysch{}.
	}
	\Description{}
	\label{fig:calb-bgs-lpt}
\end{figure}

\begin{figure}
	\centering

	\includegraphics[width=\linewidth]{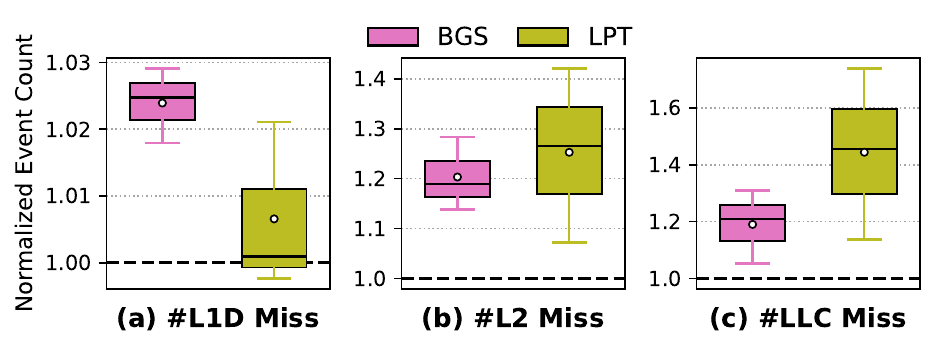}
	\caption{
		Distribution of the L1D, L2, and LLC misses incurred by the BGS and LPT scheduling strategies on G500 workloads (scale=$19$), each normalized to the corresponding \mysch{}-based execution (dashed line at $1.0$).
		Every box aggregates the samples of all three \myapp{} variants, i.e., \myapp{}-\myaccum{}, \myapp{}-HASH, and \myapp{}-ESC.
		Within each box, the solid line marks the median, the white circle the mean, and the whiskers extend to the most extreme samples within $1.5\times$ the interquartile range.
		(a) L1D misses. (b) L2 misses. (c) LLC misses.
	}
	\Description{}
	\label{fig:calb-bgs-lpt-cache-box}
\end{figure}

			We now examine the performance benefits of our \mysch{} approach. 
			Designed to alleviate cache contention while maintaining load balance, \mysch{} is evaluated against two alternative scheduling techniques also applicable to our $cf$-sorted morsels:
			(1) a bidirectional greedy scheduling scheme (BGS), where threads split into two equal-sized groups processing morsels from opposite ends toward the center, and
			(2) the classical longest-processing-time-first strategy (LPT)~\cite{DBLP:journals/siamam/Graham69}, in which all threads start from the high-$cf$ end and progress toward the low-$cf$ end.

			Figure~\ref{fig:calb-bgs-lpt} reports the throughput of BGS and LPT, normalized to our \mysch{} method, for three \myapp{}-based algorithms on G500 workloads at \textit{scale}~$19$.
			The results show that \mysch{} consistently outperforms both alternatives, achieving $7.90\%$–$19.26\%$ and $8.53\%$–$36.80\%$ higher throughput than BGS and LPT, respectively.
			While BGS also aims to mitigate contention by avoiding excessive thread concentration on high-$cf$ regions, our three-group division strategy in \mysch{} proves more effective at reducing cache pressure, especially at the high-$cf$ end.
			In contrast, LPT performs poorly precisely because it schedules all threads to handle the most demanding, high-contention tasks simultaneously, thereby exacerbating on-chip cache contention.

			This explanation rests on cache contention, so we verify it by direct measurement rather than by inference from throughput.
			Figure~\ref{fig:calb-bgs-lpt-cache-box} profiles the L1D, L2, and LLC misses incurred by BGS and LPT with hardware performance counters, again normalized to \mysch{} on the same G500 workloads, with every box aggregating the samples of all three \myapp{} variants.
			The counters are collected separately from the timing measurements and averaged over $10$ runs. Since all three policies perform the same logical SpGEMM work using the same accumulator and input, their event counts are directly comparable.
			Figure~\ref{fig:calb-bgs-lpt-cache-box}(a) shows that the three policies incur nearly identical L1D miss counts, with all measurements within $3\%$ of \mysch{}.
			This is expected because the private L1D cache is too small to hold a morsel's working set, leaving limited scope for dispatch order to improve locality at this level.
			The differences emerge in the deeper cache levels, where concurrently executing morsels compete for capacity.
			BGS and LPT incur $1.14$–$1.28\times$ and $1.07$–$1.42\times$ as many L2 misses as \mysch{}, respectively (Figure~\ref{fig:calb-bgs-lpt-cache-box}(b)).
			LPT also incurs the most LLC misses and exhibits greater variation than BGS (Figure~\ref{fig:calb-bgs-lpt-cache-box}(c)).
			These cache-miss trends are consistent with the throughput results: LPT incurs the most misses and delivers the lowest throughput, supporting cache contention as a key factor beyond load balance alone.

			This validates the rationale behind \mysch{}: by dispersing threads across the $cf$ spectrum instead of concentrating them at the high-$cf$ end, \mysch{} reduces the aggregate cache pressure of concurrently processed morsels, confirming that effective load balancing must account for cache contention.
			Notably, LPT has been a seminal scheduling strategy for over $50$ years~\cite{DBLP:journals/siamam/Graham69}, and our findings highlight the need to revisit and refine classical techniques with the idea of cache pressure alleviation.

\begin{figure}
	\centering

	\includegraphics[width=\linewidth]{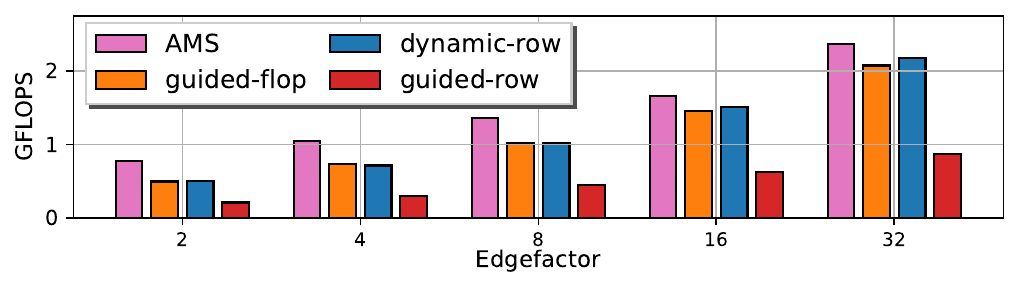}
	\vspace{-2.5em}
	\caption{
		Hashing-based SpGEMM performance on G500 matrices (scale=$17$) compared to OpenMP dynamic (-row) and guided (-row and -flop) scheduling policies.
	}
	\Description{}
	\label{fig:lb_g500}
\end{figure}
			\begin{figure}[t]
	\centering
	\includegraphics[width=\linewidth]{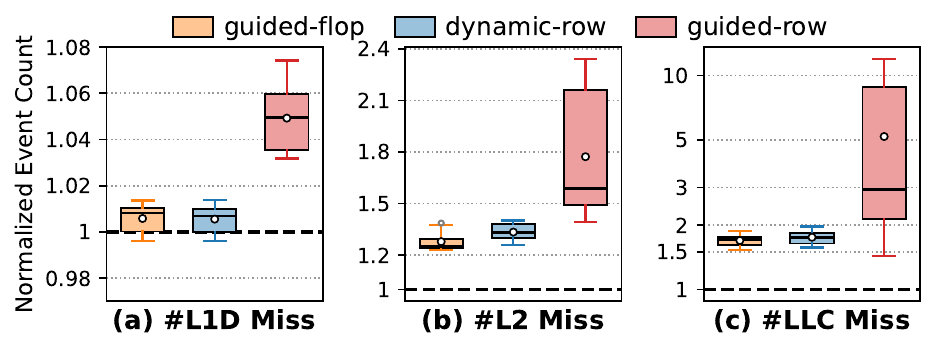}
	\caption{
		Distribution of the L1D, L2, and LLC misses incurred by the OpenMP dynamic-row, guided-flop, and guided-row scheduling policies of Figure~\ref{fig:lb_g500}, each normalized to the corresponding \myapp{}-based execution (dashed line at $1.0$).
		Within each box, the solid line marks the median, the white circle the mean, and the whiskers extend to the most extreme samples within $1.5\times$ the interquartile range.
		Note the logarithmic y-axis in (b) and (c).
		(a) L1D misses. (b) L2 misses. (c) LLC misses.
	}
	\Description{}
	\label{fig:calb-openmp-box}
\end{figure}

			We further compare \myapp{} against two widely adopted OpenMP load-balancing strategies: dynamic and guided scheduling~\cite{openmp1997openmp,DBLP:journals/tc/PolychronopoulosK87}.
			These policies schedule a unit of work different from our morsel, so they replace \myapp{}'s scheduling as a whole, and both Figure~\ref{fig:lb_g500} and Figure~\ref{fig:calb-openmp-box} are accordingly normalized to \myapp{}.
			For a fair evaluation, we implemented dynamic scheduling using an empirically determined optimal morsel size of $128$ consecutive rows (dynamic-row).
			Guided scheduling is tested with a minimum morsel set by either $128$ rows (guided-row) or a $128$-floating-point operation (FLOP) count (guided-flop); both configurations were chosen for the purpose of preserving spatial locality.

			As Figure~\ref{fig:lb_g500} illustrates, \myapp{} achieves higher throughput across all comparisons: $11.87\%$–$48.01\%$ over dynamic-row, $12.89\%$–$47.13\%$ over guided-flop, and a substantial $2.79\times$–$3.59\times$ over guided-row.
			The two margins have different origins, and Figure~\ref{fig:calb-openmp-box}, which profiles the same three policies with hardware performance counters, distinguishes them.

			Guided-row is the outlier on both figures.
			OpenMP guided scheduling progressively reduces its chunk size, which, similar to LPT~\cite{DBLP:journals/siamam/Graham69}, degrades spatial locality during later processing stages, and the profile shows exactly that: its median rises from about $1.05\times$ at L1D to about $1.6\times$ at L2 and $2.9\times$ at the LLC, with individual measurements exceeding $10\times$ \myapp{} and by far the widest spread of the three policies.
			This margin is attributable to morsel-wise dynamic scheduling itself (\S~\ref{subsec:morsel}), which keeps consecutive rows together and thereby preserves the locality that guided-row progressively gives up; that alone is enough to outperform guided-row comfortably.
			This result underscores that spatial locality is a key performance factor, and that it must be preserved together with load balancing.

			Dynamic-row and guided-flop avoid this pitfall, as their work units keep consecutive rows together and they distribute load as effectively as \myapp{} does.
			Yet both still trail \myapp{}, and the cache profiles show where the difference arises.
			At the private L1D, where a difference in spatial locality would first become visible, both remain within roughly $1\%$ of \myapp{}.
			The policies diverge only at the levels that concurrently executing morsels share, with median counts rising to $1.27$–$1.34\times$ at L2 and to $1.7$–$1.8\times$ at the LLC, a ranking consistent with their relative throughput in Figure~\ref{fig:lb_g500}.
			Since locality is comparable across the three, the remaining source of this gap is inter-thread cache contention, which \mysch{} alleviates by co-scheduling high-$cf$ and low-$cf$ morsels.
			We therefore attribute the advantage of \myapp{} over these two policies to \mysch{}, and conclude that contention-aware scheduling is essential for effective parallel load balancing.

		\subsubsection{Effectiveness of \mydaha{}}

\begin{figure}
	\centering

	\includegraphics[width=\linewidth]{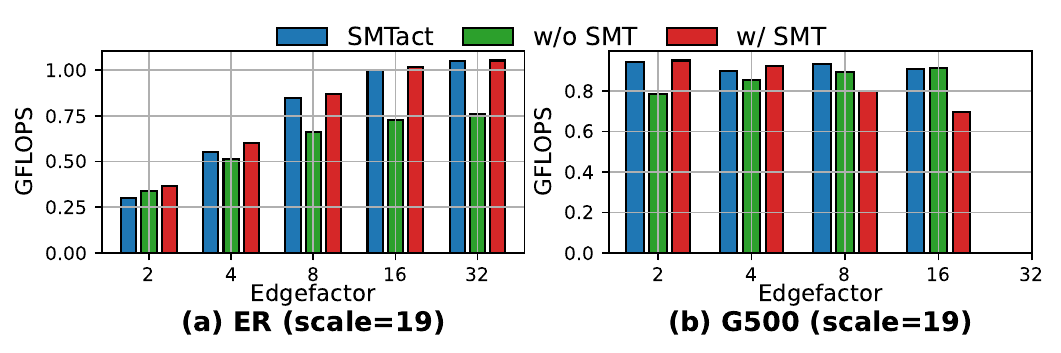}
	\vspace{-2.5em}
	\caption{
		Effectiveness of \mydaha{} in managing SMT to improve \myapp{}-\myaccum{} performance on non-skewed (ER) and skewed (G500) workloads.
	}
	\Description{}
	\label{fig:daha-cmp}
\end{figure}

			Our \mydaha{} is designed as an adaptive SMT-enablement strategy by balancing the trade-off between increased computational capability and elevated cache pressure.
			To evaluate its effectiveness, we compare \mydaha{} against two fixed configurations: SMT always enabled (w/ SMT) and SMT always disabled (w/o SMT).
			The results, presented in Figure~\ref{fig:daha-cmp}, include both non-skewed and skewed workloads.
			As illustrated, neither fixed configuration is uniformly viable: w/ SMT falls up to $23.9\%$ short of the best configuration on the skewed G500 workloads, where the extra logical threads aggravate cache contention, whereas w/o SMT forfeits up to $28.6\%$ on the non-skewed ER workloads that profit from the additional thread-level parallelism.

			Since \mydaha{} ultimately executes as one of these two configurations, its goal is not to surpass both in every case, but to automatically select a configuration at or close to the best.
			In our experiments, \mydaha{} stays within $8.2\%$ of the best fixed configuration in eight of the nine cases.
			The exception is the extremely sparse ER matrix at \textit{edgefactor}~$2$, where \mydaha{} falls $17.1\%$ short of the best fixed configuration because the sorting overhead introduced by \mysch{} is not yet amortized; this penalty is quickly offset as \textit{edgefactor} or matrix scale increases.
			Given the wide variability in sparsity patterns in real-world SpGEMM workloads, \mydaha{} thus offers a robust default that avoids the substantial degradation that either fixed policy incurs outside its sweet spot.

	\subsection{Evaluation on Bandwidth-Expanded Systems}
	\label{subsec:cxl}
		We now evaluate \myapp{} on a bandwidth-expanded memory system, where CXL boosts aggregate bandwidth but introduces higher access latency.
		As discussed in \S~\ref{sec:intro}, load imbalance can inadvertently amplify CXL's latency penalty on straggler threads, negating the bandwidth gain and making effective load balancing particularly critical.
		For this evaluation, the machine has the same CPU and OS configuration as our previous platform but features host DRAM of $28.50$~GB/s bandwidth.
		Additionally, a CXL expansion memory of $20.95$~GB/s bandwidth is installed, yielding a combined system bandwidth of $46.50$~GB/s~\cite{DBLP:conf/vldb/Huang0LCHT24,DBLP:conf/vldb/LebedevNUJA25,weisgut2025cxl}.

\begin{figure}
	\centering

	\includegraphics[width=\linewidth]{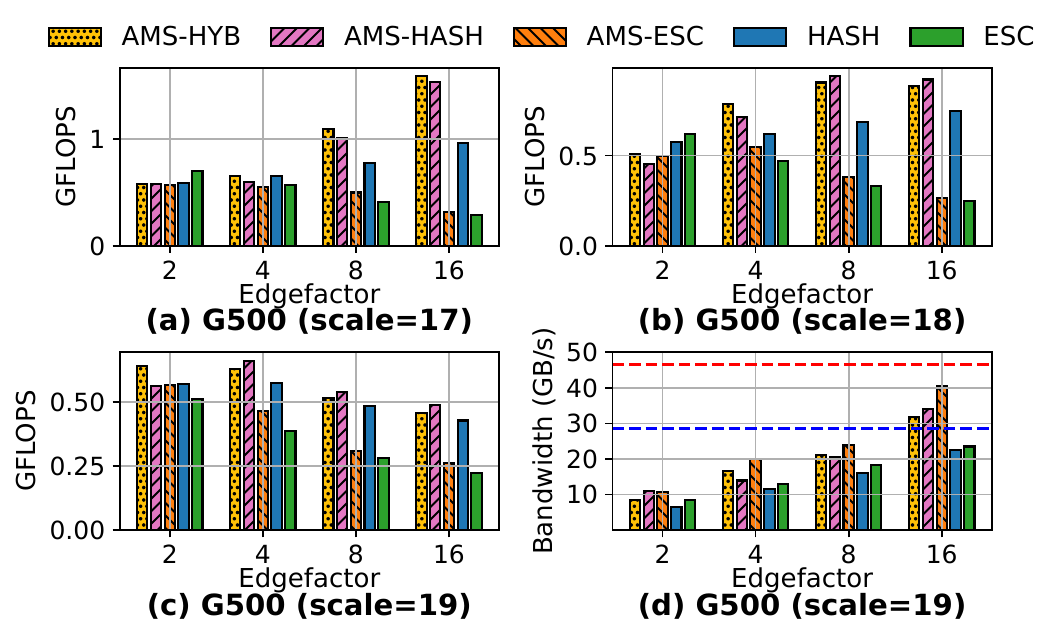}
	\vspace{-2.5em}
	\caption{
		Performance of SpGEMM on G500 workloads. 
		(a)–(c): Measured throughput at G500 scales $17$, $18$, and $19$.
		(d): Sustained bandwidth measured for different algorithms for the scale-19 workload (the red and blue dashed lines indicate the peak bandwidth of the CXL-augmented memory system ($46.50$ GB/s) and host DRAM ($28.50$ GB/s), respectively).
	}
	\Description{}
	\label{fig:g500-bw}
\end{figure}

		We evaluate \myapp{} approaches against their vanilla counterparts using skewed (G500) workloads and report the results in Figure~\ref{fig:g500-bw}.
		As shown, \myapp{}-based methods generally outperform the other methods (Figure~\ref{fig:g500-bw}(a)–(c)).
		Furthermore, the profiling results confirm that \myapp{} effectively utilizes the additional CXL-augmented bandwidth: \myapp{}-ESC sustains a bandwidth of up to $40.60$~GB/s, far exceeding the DRAM-only peak ($28.50$~GB/s) (Figure~\ref{fig:g500-bw}(d)).
		In contrast, the bandwidth consumption of the vanilla counterparts does not exceed that of the host DRAM, indicating that \myapp{} indeed improves bandwidth utilization by addressing load imbalance.

		We also observe an interesting trend in Figure~\ref{fig:g500-bw}(a)–(c): while \myapp{}-based methods remain competitive on skewed workloads, the performance gap between \myapp{} and the vanilla baselines differs for G500 matrices at \textit{scale}~$17$ versus those at \textit{scale}s~$18$ and~$19$.
		Taking the HASH-based variants as an example, \myapp{}-HASH outperforms HASH by $58.77\%$ on the G500 matrix at \textit{scale}~$17$, delivering $1.53$~GFLOPS versus $0.97$~GFLOPS (\textit{edgefactor}~$16$).
		However, this advantage diminishes to $23.16\%$ at \textit{scale}~$18$, with a similar trend observed at \textit{scale}~$19$.

		This reduction in the performance gap is mainly attributed to the coincidental load balance occasionally achieved by the static flop-count-based scheme.
		It is the extent of this imbalance that largely determines the performance degradation of baselines relative to their \myapp{} counterparts.
		As Figure~\ref{fig:hash-numc-comp-write} illustrates, computation and materialization times can exhibit opposing trends across threads.
		This is precisely the multi-dimensional skew identified in \S~\ref{subsec:skew}, namely that computation and materialization costs trend in opposite directions, and that any single-metric scheme can only coincidentally balance them.
		In some cases, these opposing patterns may randomly offset each other, resulting in a seemingly balanced overall execution time despite the underlying skew.
		This balance, however, is merely incidental and heavily dependent on the platform's relative sensitivity to computation versus memory access.
		Consequently, it fails to provide consistent or portable load balance across different architectures.

		Previous results on a $179.50$~GB/s bandwidth memory corroborate this explanation (see Figure~\ref{fig:rmat-tamu}(g)–(i)).
		With this higher bandwidth, \myapp{}-HASH achieves a $45.44\%$ performance increase over HASH ($1.52$~GFLOPS vs.\ $1.05$~GFLOPS) on the G500 matrix at \textit{scale}~$18$, because the increased bandwidth significantly reduces the platform's vulnerability to materialization cost.
		Collectively, these results validate our claim that \myapp{} provides a robust and portable solution to load balancing across diverse systems.

	\subsection{Evaluation on Other Sparse Matrix Kernels}
	\label{subsec:perf-spmmv}

\begin{figure*}
	\centering

	\includegraphics[width=\linewidth]{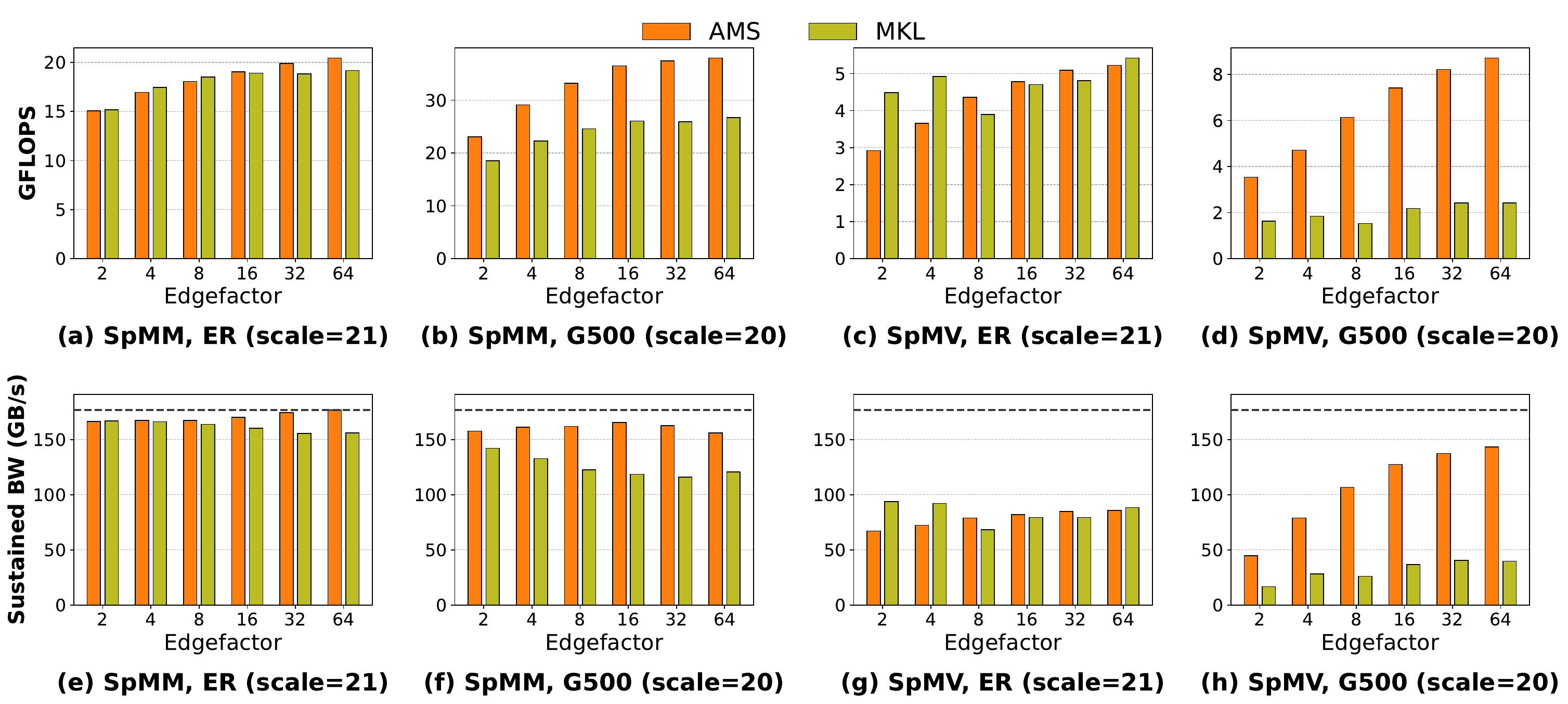}
	\caption{
		Performance of \myapp{} versus MKL for SpMM and SpMV operations on ER and G500 workloads, reported as throughput in (a)--(d) and as sustained memory bandwidth in (e)--(h).
		The dashed line marks the $179.50$~GB/s physical bandwidth of our platform's local DRAM.
	}
	\Description{}
	\label{fig:spmmmv}
\end{figure*}

		Having established \myapp{}'s portability across memory systems, we now turn to its generalizability across kernels.
		Beyond SpGEMM, we evaluate \myapp{} on other sparse matrix multiplication kernels, specifically, sparse matrix-dense matrix multiplication (SpMM) and sparse matrix-vector multiplication (SpMV).
		Since these kernels produce fixed-size dense output, we form morsels by grouping consecutive rows based on a targeted $nnz$ count in the left-hand side matrix sub-block (see \S~\ref{subsec:ext}), with each morsel sized at half the L2 cache capacity.
		The resulting morsels have disparate densities, enabling \mysch{} and \mydaha{} to alleviate cache contention as in SpGEMM.

		We compare this \myapp{}-based SpMM and SpMV implementation against the state-of-the-art Intel MKL library on both non-skewed ER and skewed G500 workloads, running on the same main-memory system as our default platform (\S~\ref{subsec:setup}), taking ER at \textit{scale}~$21$ and G500 at \textit{scale}~$20$ and sweeping the \textit{edgefactor} from $2$ to $64$, so that the behavior at higher densities remains visible.

		As shown in Figure~\ref{fig:spmmmv}(a)--(d), our method achieves on-par performance with MKL on non-skewed ER workloads while significantly outperforming it on skewed G500 workloads, with throughput gains of up to $57.25\%$ and $3.51\times$ for SpMM and SpMV, respectively.
		Figure~\ref{fig:spmmmv}(e)--(h) reports the sustained memory bandwidth of the same runs and shows where that gain originates: on the non-skewed ER workloads the two methods sustain comparable bandwidth, with SpMM approaching the $179.50$~GB/s physical limit of local DRAM, whereas on the skewed G500 workloads \myapp{} sustains substantially higher bandwidth than MKL at every \textit{edgefactor}.
		This confirms that the advantage comes from the load balancing of \myapp{} together with the on-chip cache contention it alleviates, exactly as in SpGEMM, where stragglers induced by skew leave bandwidth idle and concurrently executing dense morsels evict each other's working sets.
		By removing both, \myapp{} converts bandwidth that the baseline wastes into useful work, which is what drives the throughput gain on these memory-bound kernels.

		The bandwidth results also reveal a notable difference between SpMM and SpMV.
		SpMM sustains markedly higher bandwidth than SpMV throughout, as its dense right-hand operand offers a reuse that a single vector cannot, so SpMV wastes a far larger share of the available bandwidth, and the waste is worst under skew, where MKL sustains only a small fraction of the DRAM peak (Figure~\ref{fig:spmmmv}(h)).
		\myapp{} substantially improves bandwidth utilization on skewed SpMV workloads, sustaining roughly three times the bandwidth of MKL. At higher \textit{edgefactors}, its sustained bandwidth approaches that of SpMM.
		These results show that skew-aware scheduling improves the performance of memory-bound kernels by enabling more effective use of the available memory bandwidth.
		Together, these results demonstrate the broad applicability and robust advantages of \myapp{}'s core scheduling principles across various sparse matrix multiplication kernels.

	\subsection{Join-Aggregate and Join-Project Query Evaluation}
	\label{subsec:perf-joinagg}

		We now validate \myapp{} on two representative relational query classes: join-aggregate and join-project, using the same datasets for both.
		These queries apply aggregation or projection after a join and are widely used in prior studies of matrix multiplication for relational query processing~\cite{deep2020fast,huang2022density,DBLP:conf/icde/SunKH23}.
		By modeling tables as sparse matrices, SpGEMM fuses the join with the subsequent aggregation or projection, avoiding materialization of the intermediate join result~\cite{amossen2009faster,DBLP:journals/pacmmod/KhamisHS25,DBLP:conf/icde/AbergerLOR18}.
		This is particularly beneficial for many-to-many joins and self-joins that produce large intermediate results.

		We employ widely used datasets from prior works~\cite{huang2022density,deep2020fast,DBLP:conf/icde/SunKH23,DBLP:conf/sigir/HuangL0FY20}, and also evaluate on an industrial OLAP ClickHouse benchmark~\cite{DBLP:journals/pvldb/SchulzeSYDM24} with two sampled workloads: CLH0, a uniform $5\%$ sample of users, and CLH1, the top $5\%$ of users by pageviews (see Table~\ref{tab:groupby-flop} for key statistics).
		These datasets span a wide range of skewness, from mildly to highly skewed, as quantified by the Gini coefficient in Table~\ref{tab:groupby-flop}.
		Notably, the uniform sample exhibits a higher Gini coefficient than the top-user sample ($0.59$ versus $0.39$), because selecting only the heaviest users excludes the low-activity tail, whereas uniform sampling largely preserves the original population’s skew.
		As for the query form, we perform self-join (i.e., $R=S$) with aggregation in the form:
		\begin{small}
			\begin{verbatim}
				SELECT R.user, S.user, SUM(R.val * S.val)
				FROM R, S
				WHERE R.item = S.item AND R.user <> S.user
				GROUP BY R.user, S.user
			\end{verbatim}
		\end{small}
		The corresponding join-project query replaces the aggregation with a projection over the same self-join:
		\begin{small}
			\begin{verbatim}
				SELECT DISTINCT R.user, S.user
				FROM R, S
				WHERE R.item = S.item AND R.user <> S.user
			\end{verbatim}
		\end{small}
		\vspace{-1em}
		Both queries measure user-wise similarity, where ``item'' and ``val'' represent different attributes and interaction values depending on the scenario, e.g., URL/click-count in ClickHouse (CLH0 and CLH1) and movie/rating in MovieLens (MVLN).
		The aggregate query form is particularly challenging, as the aggregated values increase the demand for memory, and it is not amenable to existing aggregate query rewriting techniques~\cite{DBLP:conf/vldb/ChaudhuriS94,DBLP:conf/vldb/YanL95,DBLP:journals/vldb/EichFM18,DBLP:journals/vldb/FentBN23,DBLP:conf/sigmod/WangC18,DBLP:conf/icde/WangC20,DBLP:conf/icde/YanL94,DBLP:journals/pvldb/Fent021,DBLP:journals/pvldb/MoerkotteN11}.

\begin{table}[t]
\centering
\caption{Summary of the $10$ relational workloads used for both join-aggregate and join-project evaluation.}
\label{tab:groupby-flop}
\resizebox{\columnwidth}{!}{%
\begin{tabular}{lrrrrrr}
	\toprule
	\textbf{Dataset} & \textbf{$|user|$} & \textbf{$|tuple|$} & \textbf{Gini} & \textbf{$|OUT_J|$} & \textbf{$|OUT_A|$} & \textbf{$|OUT_J|/|OUT_A|$} \\
	\midrule

	IMDB~\cite{imdb2024datasets} 	& 733.47K	& 7.52M	& 0.3453	& 214.77M	& 176.91M	& 1.2140	\\
	ISF18~\cite{DBLP:journals/isf/NguyenHTK18} 	& 1.82K	& 1.03M	& 0.5586	& 327.58M	& 3.29M	& 99.6586	\\
	CLH0~\cite{DBLP:journals/pvldb/SchulzeSYDM24} 	& 850.00K	& 2.90M	& 0.5858	& 2.43B	& 2.34B	& 1.0415	\\	%
	RD23~\cite{DBLP:conf/um/SunK0K23} 	& 6.72K	& 3.91M	& 0.5276	& 4.61B	& 45.02M	& 102.3613	\\
	BKGN~\cite{kotkov2022tag} 	& 350.33K	& 5.15M	& 0.7242	& 11.43B	& 6.77B	& 1.6892	\\
	GDBK~\cite{gdbk} 	& 53.42K	& 5.98M	& 0.1293	& 19.63B	& 2.51B	& 7.8261	\\
	CLH1~\cite{DBLP:journals/pvldb/SchulzeSYDM24} 	& 850.00K	& 24.05M	& 0.3862	& 20.34B	& 11.41B	& 1.7823	\\	%
	JOKE~\cite{huang2022density} 	& 23.50K	& 1.71M	& 0.1709	& 32.65B	& 552.25M	& 59.1219	\\	%
	SAC18~\cite{DBLP:conf/sac/KotkovKZV18} 	& 104.66K	& 10.00M	& 0.6988	& 68.52B	& 5.94B	& 11.5325	\\
	MVLN~\cite{huang2022density} 	& 138.49K	& 20.00M	& 0.5807	& 269.61B	& 16.80B	& 16.0477	\\

	\bottomrule
	\multicolumn{7}{p{\linewidth}}{%
		\smallskip
		\footnotesize
		\begin{enumerate}[leftmargin=*, nosep, label=\arabic*.]
			\item $|user|$ and $|tuple|$ are equivalent to the matrix row count and total nnz count, respectively, reflecting the relational-to-sparse-matrix duality; Gini captures the user-wise nnz distribution~\cite{DBLP:conf/websci/KunegisP12}.
			\item $|OUT_J|$ and $|OUT_A|$ represent the cardinalities of the join result and the final aggregated result, respectively; the projected result has the same cardinality.
		\end{enumerate}
	} \\
\end{tabular}
}
\end{table}

		\begin{table}[t]
	\centering
	\caption[Summary of all evaluated join-aggregate query processing approaches]{%
		Summary of all evaluated join-aggregate query processing approaches}
	\label{tab:ja-methods}
	\renewcommand{\arraystretch}{1.2}
	\setlength{\tabcolsep}{2pt} 
	\footnotesize
	
	\begin{tabularx}{\columnwidth}{@{} >{\ttfamily\raggedright\arraybackslash}p{1.2cm} >{\centering\arraybackslash}p{1.6cm} X @{}}
		\toprule
		\normalfont\textbf{Method}
			& \textbf{In-place Agg.\textsuperscript{$*$}}
			& \textbf{Description} \\
		\midrule
		\rowcolor{groupbg}
		\multicolumn{3}{l}{\textit{Traditional join followed by aggregation}} \\
		PHJ-AGG~\cite{DBLP:conf/sigmod/SchuhCD16,skarupke2017fastest}
			& \xmark
			& Radix partitioned hash join with aggregation; deduplication via fast hash table~\cite{huang2022density,skarupke2017fastest}. \\

		NPHJ-AGG~\cite{DBLP:conf/sigmod/BlanasLP11,skarupke2017fastest}
			& \cmark
			& Non-partitioned hash join; aggregates computed during probing to skip the final deduplication pass~\cite{huang2022density,skarupke2017fastest}. \\
		\midrule
		\rowcolor{groupbg}
		\multicolumn{3}{l}{\textit{Join-alone algorithms (no aggregation; for comparison only)}} \\
		PHJ~\cite{DBLP:conf/sigmod/SchuhCD16}
			& {---}
			& Radix partitioned hash join; serves as the join backend of PHJ-AGG. Does not produce aggregation results. \\

		NPHJ~\cite{DBLP:conf/sigmod/BlanasLP11}
			& {---}
			& Non-partitioned hash join; serves as the join backend of NPHJ-AGG. Does not produce aggregation results. \\
		\midrule
		\rowcolor{groupbg}
		\multicolumn{3}{l}{\textit{State-of-the-art join-aggregate techniques}} \\
		DIM3~\cite{huang2022density}
			& \cmark
			& Hybrid sparse/dense join-aggregate query algorithm; intersection-free partitioning eliminates deduplication. \\
		\midrule
		\rowcolor{groupbg}
		\multicolumn{3}{l}{\textit{Industry SpGEMM competitors}} \\
		MKL~\cite{intel-mkl}
			& \cmark
			& Intel oneMKL SpGEMM (non-sorted variant); aggregation replaces deduplication inside the accumulator. \\
		\midrule
		\rowcolor{groupbg}
		\multicolumn{3}{l}{\textit{Proposed \myapp{}-based approaches}} \\
		\rowcolor{ablbg}
		\myapp{}-HYB
			& \bm{\cmark}
			& \myapp{} scheduling with hybrid HASH/ESC selection; eliminates the separate deduplication pass. \\
		\bottomrule
	\end{tabularx}
	\vspace{2pt}
	\begin{minipage}{\linewidth}
		\scriptsize
		\textsuperscript{$*$} \emph{In-place Agg.} denotes whether aggregation is performed on-the-fly alongside the join, or both operations are directly merged into one.
	\end{minipage}
\end{table}

\begin{figure*}
	\centering

	\includegraphics[width=\linewidth]{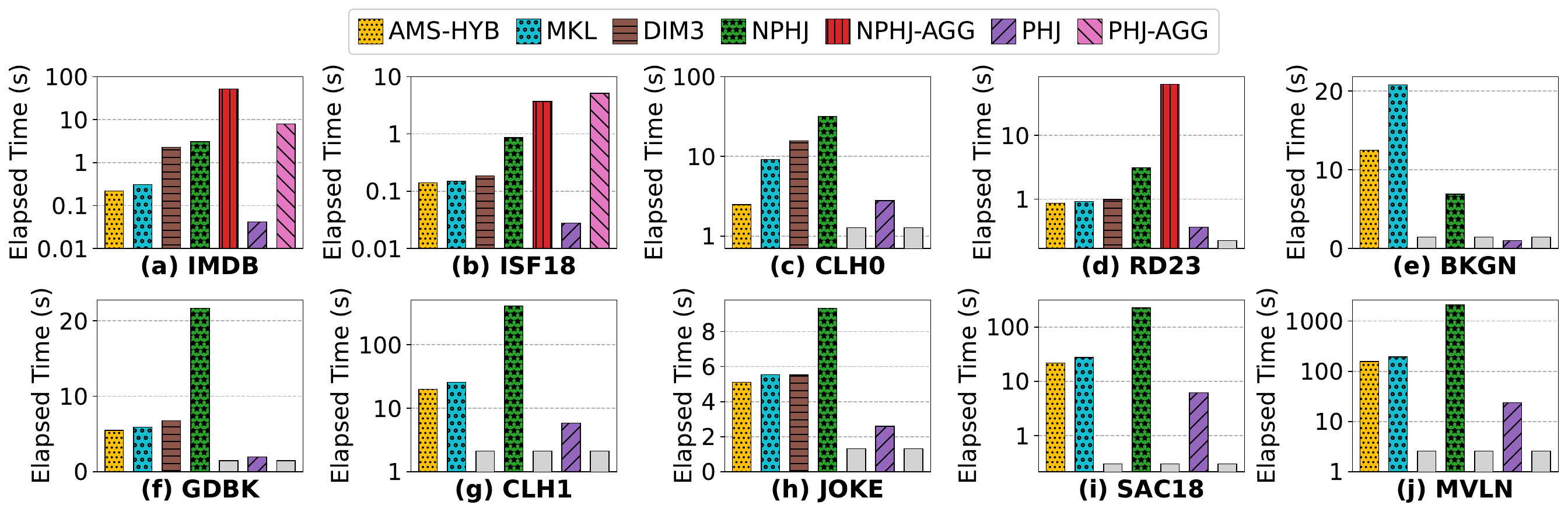}

	\caption{
		Execution time comparison across diverse join-aggregate workloads;
		gray bars with ``N/A'' denote algorithms that run out of memory on those specific workloads;
		note that vertical axes use a logarithmic scale to capture the wide range of execution times, except for subfigures (e), (f), and (h).
	}
	\Description{}
	\label{fig:groupby-eva}
\end{figure*}

		Relational join queries do not require ordered output by default, and sorting the output within each row adds substantial overhead, as shown in \S~\ref{subsec:perf} and reported in prior SpGEMM studies~\cite{DBLP:conf/icppw/NagasakaMAB18,DBLP:journals/pc/NagasakaMAB19}.
		We therefore conduct this evaluation in the non-sorted regime, comparing \myapp{}-\myaccum{} with: (1) DIM3~\cite{huang2022density}, the state-of-the-art join-aggregate method based on SpGEMM; (2) NPHJ-AGG, where a fast hash table~\cite{skarupke2017fastest} performs aggregation on-the-fly alongside the probing phase of non-partitioned hash join (NPHJ)~\cite{DBLP:conf/sigmod/SchuhCD16,DBLP:conf/sigmod/BandleG021,DBLP:journals/pvldb/HuangJZHT23}, also bypassing intermediate materialization; (3) PHJ-AGG, where we use the cache-efficient partitioned hash join~\cite{DBLP:conf/icde/BalkesenTAO13,DBLP:journals/tkde/BalkesenTAO15,DBLP:journals/pvldb/BalkesenATO13,DBLP:conf/vldb/BonczMK99} to perform the join and materialize the intermediate result, and after that we use the same efficient hash table~\cite{skarupke2017fastest} as NPHJ-AGG to perform the aggregation
		~\footnote{We do not have PHJ-AGG perform the on-the-fly aggregation as NPHJ-AGG, because on-the-fly aggregation would incur substantial memory traffic to the aggregation hash table, breaking the cache efficiency that partitioning is designed to deliver~\cite{DBLP:journals/tkde/ManegoldBK02,DBLP:conf/icde/BalkesenTAO13,DBLP:journals/tkde/BalkesenTAO15}.}
		; (4) the industrial SpGEMM method: MKL with non-sorted output.
		We refer to NPHJ-AGG and PHJ-AGG collectively as the ``join-then-aggregate'' methods, as both evaluate the query in a traditional way, performing the aggregation on top of an explicit join rather than fusing the two as the matrix-based methods do.
		For reference, we also include NPHJ and PHJ as join-only baselines (which do not materialize the intermediate results).
		All evaluated methods are summarized in Table~\ref{tab:ja-methods} for a clear reference.

		\begin{table}[t]
	\centering
	\caption{
		Tuple-to-CSR conversion cost, covering both operands of each query,
		and its share of the total \myapp{}-\myaccum{} runtime under the two query classes.
	}
	\label{tab:tuple2csr}
	\normalsize
	\begin{tabular*}{\columnwidth}{@{\extracolsep{\fill}}lrrr@{}}
		\toprule
		\multirow{2}{*}{\textbf{Dataset}} & \multirow{2}{*}{\textbf{Conversion (s)}} & \multicolumn{2}{c}{\textbf{Share (\%)}} \\
		\cmidrule(l){3-4}
		& & \textbf{Join-Aggregate} & \textbf{Join-Project} \\
		\midrule
		IMDB    & 0.0126 & 5.7536 & 7.3995 \\
		ISF18   & 0.0043 & 3.0764 & 4.1958 \\
		CLH0    & 0.0087 & 0.3534 & 0.4729 \\
		RD23    & 0.0064 & 0.7476 & 1.1090 \\
		BKGN    & 0.0128 & 0.1025 & 0.1260 \\
		GDBK    & 0.0097 & 0.1764 & 0.3282 \\
		CLH1    & 0.0679 & 0.3391 & 0.6139 \\
		JOKE    & 0.0049 & 0.0962 & 0.1948 \\
		SAC18   & 0.0156 & 0.0720 & 0.1422 \\
		MVLN    & 0.0488 & 0.0313 & 0.0807 \\
		\bottomrule
	\end{tabular*}
\end{table}

\begin{figure*}
	\centering

	\includegraphics[width=\linewidth]{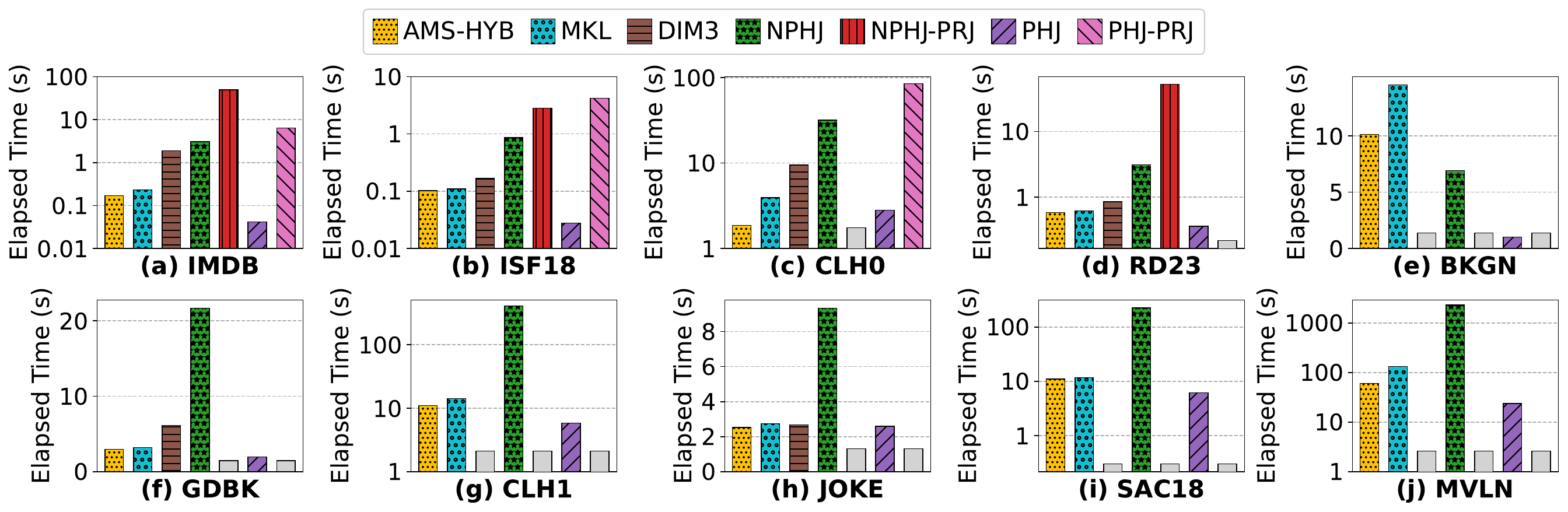}

	\caption{
		Execution time comparison across diverse join-project workloads;
		gray bars with ``N/A'' denote algorithms that run out of memory on those specific workloads;
		note that vertical axes use a logarithmic scale to capture the wide range of execution times, except for subfigures (e), (f), and (h).
	}
	\Description{}
	\label{fig:groupby-jp-eva}
\end{figure*}

		As for the join-project setting, we use the same set of methods but replace the aggregation with a projection.
		A projection no longer references the value domain, so both the matrix-based and the traditional ``join-then-project'' methods can omit the value computation altogether, avoiding the memory traffic and arithmetic cost associated with processing these values, and every method that produces the query output is correspondingly faster than in the join-aggregate setting above.
		For the matrix-based methods, i.e., \myapp{}-\myaccum{}, MKL, and DIM3, the value array of the output CSR matrix is thus neither allocated nor populated, and the projection is read directly from the sparsity structure of the matrix product.
		Moreover, join-project corresponds to a Boolean matrix product, so only the existence of each output nonzero matters.
		Following prior work~\cite{amossen2009faster,huang2022density}, we therefore enable early termination in \myapp{}-\myaccum{} and DIM3, stopping the accumulation for each output cell as soon as it is known to be nonzero.
		In contrast, join-aggregate must sum all contributions to each output cell.
		It should be noted that we are not able to apply this mechanism to MKL, as the proprietary library exposes only a numerical product.
		For the traditional ``join-then-project'' methods, denoted NPHJ-PRJ and PHJ-PRJ, no value payload is carried or combined, and the two mirror their aggregate counterparts: NPHJ-PRJ projects on-the-fly alongside probing as NPHJ-AGG does, whereas PHJ-PRJ projects only after the intermediate result is materialized, for the same reason as PHJ-AGG.

		It is noteworthy to mention that in this evaluation, the table inputs are in the original row-wise tuple format $\langle$row-index, column-index, value$\rangle$, e.g., $\langle$user, URL, click count$\rangle$ for the ClickHouse workloads, which is equivalent to the coordinate format (COO) in sparse matrix storage~\cite{DBLP:conf/ipps/FlynnYSKY25,saad2003iterative,DBLP:journals/pacmmod/KimR26}%
		~\footnote{For join-project, only existence matters, so a recorded pair is the nonzero entry itself and the row-wise tuple can drop the value, leaving $\langle$row-index, column-index$\rangle$ alone as exactly the COO of a sparse Boolean matrix.}%
		.
		The hash-join methods operate on these tuples directly, whereas the matrix-based methods require CSR and therefore incur a preliminary pass of data format conversion.
		We perform this tuple-to-CSR conversion with a parallel pass of histogram, prefix sum, and scatter, following the common practice of mainstream sparse-matrix libraries~\cite{scipy2020,eigen}, and charge its time to all matrix-based methods.
		Table~\ref{tab:tuple2csr} reports this cost on every workload, covering both operands of each query.
		The conversion costs at most $7.40\%$ of the \myapp{}-\myaccum{} runtime, with the two largest shares on IMDB and ISF18, which have the lowest $|OUT_J|/|tuple|$ ratios and therefore the least multiplication over which to amortize the linear pass.
		On the remaining eight workloads it stays below $1.2\%$ under both query classes, so the input format is not a performance concern for the matrix-based methods.

		Figure~\ref{fig:groupby-eva} presents the join-aggregate results. 
		As shown, \myapp{}-\myaccum{} consistently achieves the best performance among all aggregate variants, demonstrating its effectiveness in addressing relational join queries.
		Surprisingly, even compared with the join-only baseline NPHJ, \myapp{}-\myaccum{} wins $9$ out of $10$ cases, losing only on BKGN, highlighting the superiority of SpGEMM methods in handling joins with large output cardinalities.
		Comparing the two runners-up, MKL and DIM3, DIM3 is the weaker of the two, trailing MKL by $1.65\times$ in geometric mean over the workloads both complete, and it fails to finish on some workloads due to out-of-memory errors.
		This is because DIM3 uses an ``estimate-partition-multiplication'' approach that tends to over-estimate output size under skew, a common scenario in real-world workloads, and thus incurs significant memory materialization cost.
		Moreover, DIM3 and other prior matrix-based works~\cite{deep2020fast,hu2020parallel,hu2025output,DBLP:conf/icde/SunKH23} mainly target an algorithmic remapping, i.e., expressing the query as a sparse matrix product, and leave the parallelization to a plain OpenMP loop without a scheduling design of their own.
		DIM3 therefore inherits exactly the load imbalance that \myapp{} targets, which is why it generally trails MKL, an industrial standard with mature thread scalability, even where DIM3 completes.
		Notably, \myapp{}-\myaccum{} outperforms MKL by $1.37\times$ in geometric mean and by up to $3.69\times$ on CLH0, a skewed workload with a sizable matrix of $850$K rows and a Gini coefficient of $0.59$.
		These results show that \myapp{}'s benefits from addressing multi-dimensional skew and reducing on-chip cache contention translate directly to skewed relational workloads.
		These gains are most pronounced on large, skewed workloads, where load imbalance and cache contention can make memory accesses dominate runtime.
		Smaller workloads, by contrast, fit largely in cache, leaving less memory-access overhead for \myapp{} to reduce.

		The hash-join baselines face more severe limitations on larger workloads: of the ten workloads, NPHJ-AGG completes only the three smallest, while PHJ-AGG completes only the two smallest.
		This is because, regardless of whether the aggregation is performed on-the-fly or after the join result is fully populated, the hash table must grow dynamically to hold the unhashed entries that are to be aggregated.
		This dynamically growing hash table is usually resized with a specific scaling factor (typically $2\times$), and therefore exhausts the memory footprint very quickly.
		Beyond memory limitations, comparing the ``join-then-aggregate'' approaches with their respective join-only baselines, we can see that the aggregation-based methods are more than an order of magnitude slower on average, indicating the significant cost of aggregation, and thus underscoring the necessity and importance of SpGEMM-based methods.
		This is particularly true for PHJ and PHJ-AGG: PHJ alone achieves the best runtime on all workloads but CLH0, yet aggregation inflates its runtime by more than two orders of magnitude wherever PHJ-AGG completes (for instance from $0.041$ to $7.852$ seconds on IMDB).
		This stark contrast indicates that the aggregation operation itself is the dominant bottleneck, underscoring the need for effective join-aggregate query processing.

		We now turn to the join-project setting, which shares the same joins, datasets, and skew.
		Figure~\ref{fig:groupby-jp-eva} reports this second setting, and the performance ranking of the methods carries over unchanged.
		\myapp{}-\myaccum{} again achieves the best performance among all join-project variants.
		MKL also follows as the closest competitor, trailing by $1.33\times$ in geometric mean and by up to $2.19\times$ on MVLN.
		DIM3 again ranks third, but its gap behind \myapp{}-\myaccum{} widens from $2.28\times$ in join-aggregate to $2.64\times$ in join-project, measured in geometric mean over the workloads it completes.
		It also continues to fail on the same four workloads with the largest outputs, for the reasons discussed previously.

		Compared with the join-aggregate results, the join-project workloads generally exhibit lower runtime, as the query no longer requires the value computation and thereby incurs less arithmetic cost and less memory traffic.
		\myapp{}-\myaccum{}, DIM3, and MKL accordingly run $1.66\times$, $1.34\times$, and $1.71\times$ faster in geometric mean than their respective join-aggregate counterparts.
		DIM3, however, still falls behind MKL by $2.10\times$ in geometric mean over the workloads both complete, losing on five of the six, even though DIM3 uses early termination and MKL does not.
		This underscores that, without proper scheduling and load balancing, performing less work does not always translate into higher performance.
		\myapp{}-\myaccum{}, in contrast, pairs early termination with contention-aware scheduling and load balancing, and therefore converts the reduced work into an actual runtime saving.

		The hash-join baselines face similar limitations to their join-aggregate counterparts, with NPHJ-PRJ and PHJ-PRJ each completing only three of the ten workloads.
		This is because both require a dynamically growing hash table for projection, whether it is performed on-the-fly or after the join result is fully materialized.
		As the table grows, it typically doubles in size with each resize, eventually exhausting memory on larger workloads.
		Beyond these memory limitations, NPHJ-PRJ remains up to $17\times$ slower than the join-only NPHJ on workloads it completes, showing that projection dominates the runtime.
		\myapp{}-\myaccum{} again wins $9$ of the $10$ cases against NPHJ itself, losing only on BKGN, which reaffirms the superiority of SpGEMM-based methods on relational join-project queries.

	\subsection{Graph-Analytic Evaluation}
	\label{subsec:perf-graph}

		Real-world graph workloads are typically sparse, and are commonly modeled as sparse matrices through their adjacency representation, which places one nonzero at every edge, and in this representation, expanding a set of vertices to their neighbors can be expressed as a matrix product.
		SpGEMM is therefore an important building block of graph analytics, expressing operators such as frontier expansion, neighborhood composition, and masked edge enumeration.
		Moreover, many real-world graphs have power-law degree distributions, which lead to skewed row lengths in their adjacency matrices.
		This makes them well suited to \myapp{}'s contention-aware scheduling and load balancing.
		We evaluate three representative graph-analytic workloads, namely meta-path construction for PathSim similarity search, breadth-first search, and triangle counting.

		\subsubsection{PathSim Meta-Path Construction}
		\label{subsec:perf-pathsim}

			We first evaluate \myapp{} on PathSim~\cite{DBLP:journals/pvldb/SunHYYW11}, a meta-path-based similarity measure for heterogeneous information networks, focusing on the intermediate matrices it must construct.
			Given a symmetric meta-path, PathSim scores node pairs through a commuting matrix $M = M_l M_l^{T}$, whose left half $M_l$ is a chain of sparse matrix products.
			Constructing $M_l$ is therefore an SpGEMM workload, and it dominates the cost of answering a query.

\begin{table}[t]
\centering
\caption{$M_l$ construction on DBLP V10 (seconds).}
\label{tab:pathsim}
\resizebox{\columnwidth}{!}{%
	\begin{tabular}{lrrrrrr}
		\toprule
		\textbf{Meta-path} & \textbf{$|row|$} & \textbf{$|nnz|$} & \textbf{Gini} & \textbf{\myapp{}-\myaccum{}} & \textbf{MKL} & \textbf{Speedup} \\
		\midrule
		APCPA	& 1.77M		& 4.98M		& 0.62	& 0.052	& 0.117	& $2.24\times$	\\
		CPAPC	& 5.08K		& 4.98M		& 0.78	& 0.067	& 0.120	& $1.80\times$	\\
		\addlinespace
		APTPA	& 1.77M		& 48.48M	& 0.60	& 0.158	& 0.529	& $3.34\times$	\\
		TPAPT	& 120.69K	& 48.48M	& 0.94	& 0.235	& 0.382	& $1.62\times$	\\
		\midrule
		Geomean	& --		& --		& --	& 0.107	& 0.231	& $2.16\times$	\\
		\bottomrule
	\end{tabular}
	}
\end{table}

\begin{figure*}
	\centering
	\includegraphics[width=\linewidth]{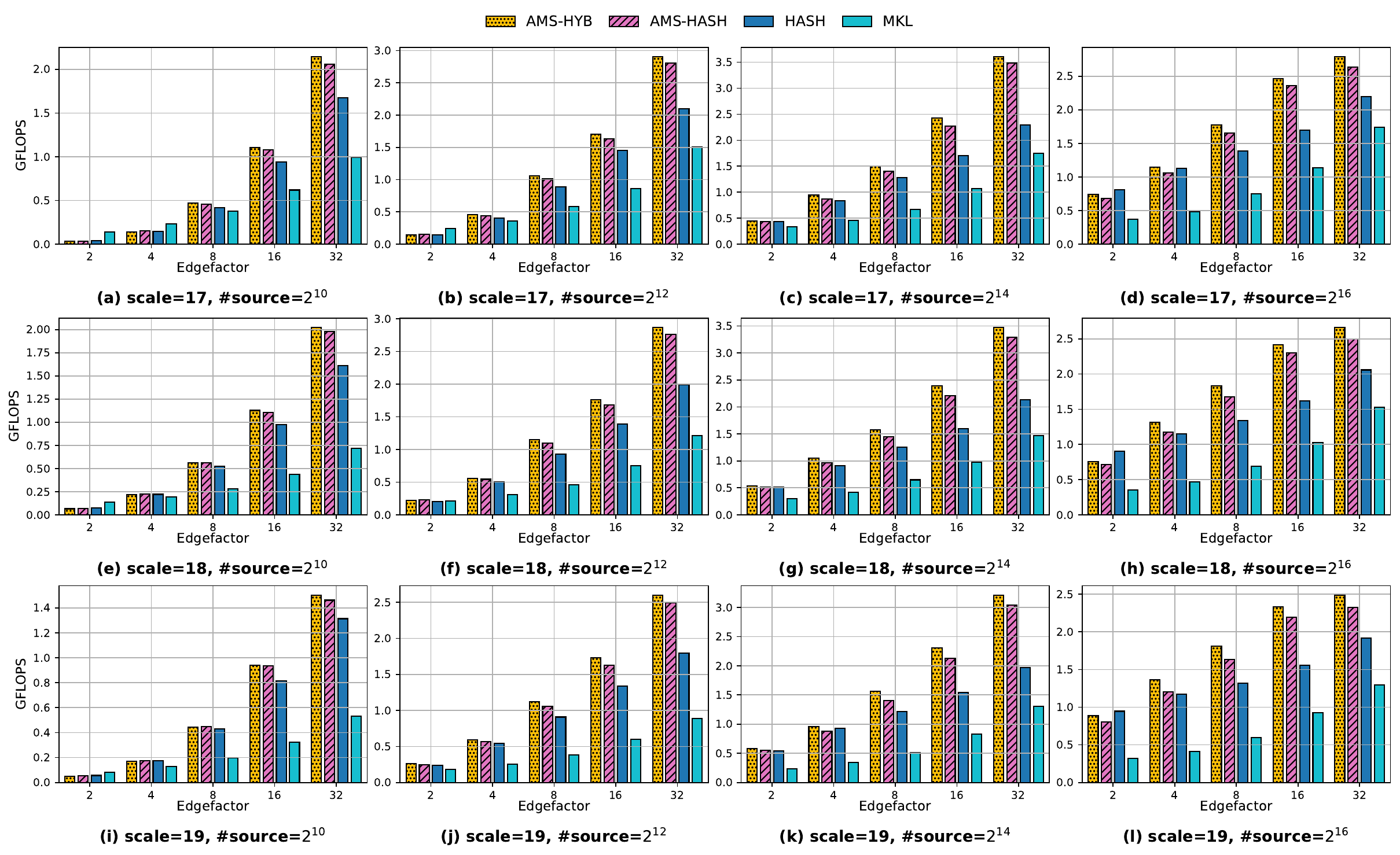}
	\caption{
		Multi-source BFS throughput on synthetic G500 graphs, where a frontier matrix holding $\#source$ sources is repeatedly multiplied by the adjacency matrix.
		Subfigures (a)--(l) sweep the graph \textit{scale} over $17$--$19$ and the number of sources over $2^{10}$--$2^{16}$, with \textit{edgefactor} varying along the x-axis of every panel.
	}
	\Description{}
	\label{fig:bfs}
\end{figure*}

			We evaluate on the AMiner DBLP V10 citation network~\cite{aminer_dblp_v10}, comprising $1.77$M authors (A), $3.08$M papers (P), $5{,}078$ venues (C), and $120{,}685$ terms (T), with $9.48$M author-paper, $2.57$M venue-paper, and $22.64$M paper-term edges.
			We select four symmetric meta-paths that place each node type in the row dimension: APCPA and APTPA relate authors through shared venues and terms, whereas TPAPT and CPAPC relate terms and venues through shared authors.
			As Table~\ref{tab:pathsim} summarizes, they differ widely in both shape and skew, and form two transpose pairs (APCPA/CPAPC and APTPA/TPAPT) that perform the same work over row counts differing by $350\times$ and $15\times$.
			All four exhibit pronounced row-wise skew (Gini $0.60$--$0.94$), reflecting prevalent skew in real-world graphs.
			We compare \myapp{}-\myaccum{} against the industry-standard MKL and report the SpGEMM time of the $M_l$ construction.

			As reported in Table~\ref{tab:pathsim}, \myapp{}-\myaccum{} outperforms MKL on all four meta-paths, by $1.62\times$ to $3.34\times$, giving a geometric-mean speedup of $2.16\times$.
			These gains span a $350\times$ range of row counts, from the $5{,}078$-row CPAPC to the $1.77$M-row APTPA, indicating that the benefit of \myapp{} does not depend on a particular matrix shape.
			Within each transpose pair, however, the gain is consistently larger for the variant with more rows ($2.24\times$ against $1.80\times$, and $3.34\times$ against $1.62\times$), as the greater computational work associated with the larger row count helps amortize the cost of $cf$-based morsel sorting.
			Overall, \myapp{} extends its advantage beyond matrix squaring and relational queries to meta-path construction on heterogeneous networks.

		\subsubsection{Breadth-First Search}
		\label{subsec:perf-bfs}

			We next evaluate \myapp{} on multi-source breadth-first search (BFS), a traversal primitive used in many graph-analytic tasks~\cite{DBLP:journals/pvldb/DhulipalaMKGBGS20}.
			For example, betweenness centrality measures how often a vertex lies on shortest paths between other vertices.
			Computing it on an unweighted graph requires a BFS from each source vertex to obtain shortest-path counts, making repeated traversals over the same graph a major part of its cost.
			Beyond centrality, BFS also supports connected components, reachability and $k$-hop neighborhood queries, unweighted single-source shortest paths, and diameter or eccentricity estimation.
			Faster traversal can therefore accelerate a broad range of graph-analytic workloads.

			SpGEMM is exactly the kernel that carries out one traversal level in the linear-algebraic formulation~\cite{DBLP:journals/pc/NagasakaMAB19}.
			The left operand is the graph's square adjacency matrix, while the right operand is a tall-skinny frontier matrix with one column per source, marking the vertices reached at the current level.
			Following the row-wise formulation in \S~\ref{subsec:spgemm}, each adjacency row selects and accumulates the frontier rows corresponding to its neighbors.
			The vertex degree therefore determines how many frontier rows must be processed, while a single matrix product advances all sources by one hop.
			Multi-source BFS thus consists of repeated square-by-tall-skinny products, whose row-wise work reflects the graph's degree distribution.
			Since real-world graphs commonly exhibit power-law degree distributions, these products naturally exhibit the row-wise skew that \myapp{} is designed to handle.

			Following the setup of Nagasaka et al.~\cite{DBLP:journals/pc/NagasakaMAB19}, we evaluate the BFS kernel as the product of a square graph adjacency matrix and a tall-skinny frontier matrix on synthetic G500 graphs.
			These graphs capture the power-law degree distributions common in real-world graphs while allowing the frontier width to vary independently of graph size and density.
			We sweep the graph \textit{scale} over $17$--$19$ and the number of concurrent sources over $2^{10}$--$2^{16}$, and extend that setup by sweeping the \textit{edgefactor} over $2$--$32$ as well, so that density varies alongside frontier width.
			As the frontier is used as a set of reached vertices, frontier expansion requires no output ordering, so we report only the non-sorted methods.
			We compare \myapp{}-\myaccum{} and \myapp{}-HASH against the HASH baseline and the non-sorted Intel MKL variant.

			Figure~\ref{fig:bfs} shows that \myapp{}-\myaccum{} achieves the highest throughput among the four methods on $47$ of the $60$ evaluated workloads.
			It outperforms the HASH baseline on $50$ workloads, with a geometric-mean speedup of $1.18\times$ and a maximum speedup of $1.63\times$, while \myapp{}-HASH achieves a geometric-mean speedup of $1.13\times$ over the same baseline.
			The advantage over MKL is larger: \myapp{}-\myaccum{} achieves a geometric-mean speedup of $1.91\times$ and a maximum speedup of $3.31\times$.

			The remaining $13$ workloads, where \myapp{}-\myaccum{} does not achieve the highest throughput, occur at the extremes of the parameter sweep.
			MKL wins on $5$ workloads with low density and few sources (\textit{edgefactor}~$2$--$4$ and $2^{10}$--$2^{12}$ sources), where the limited computation is insufficient to amortize \myapp{}'s morsel preparation cost.
			The HASH baseline wins on another $3$, all at \textit{edgefactor}~$2$ with the widest frontier ($2^{16}$ sources).
			Finally, \myapp{}-HASH wins on the remaining $5$ by margins below $5\%$, with $4$ of these occurring at $2^{10}$ sources.

			From Figure~\ref{fig:bfs}, we can observe two trends in \myapp{}-\myaccum{}'s speedup over the HASH baseline.
			First, the speedup increases steadily with graph density, from $0.96\times$ at \textit{edgefactor}~$2$ to $1.38\times$ at \textit{edgefactor}~$32$.
			Only at the lowest density does \myapp{}-\myaccum{} fall behind the baseline, because the computation is insufficient to amortize the cost of $cf$-based morsel sorting, consistent with the matrix-squaring results in \S~\ref{subsec:perf}.
			Second, the speedup increases with the number of concurrent sources, from $1.05\times$ at $2^{10}$ sources to $1.20\times$ at $2^{16}$ sources.
			A wider frontier increases the work performed at each traversal level, creating more cache contention for \mysch{} to alleviate.
			Together, these trends suggest that \myapp{} benefits most when the workload provides enough computation to amortize scheduling overhead and enough skew-induced contention for its scheduling strategy to address.

		\subsubsection{Triangle Counting}
		\label{subsec:perf-tc}

\begin{figure}[t]
	\centering
	\includegraphics[width=\linewidth]{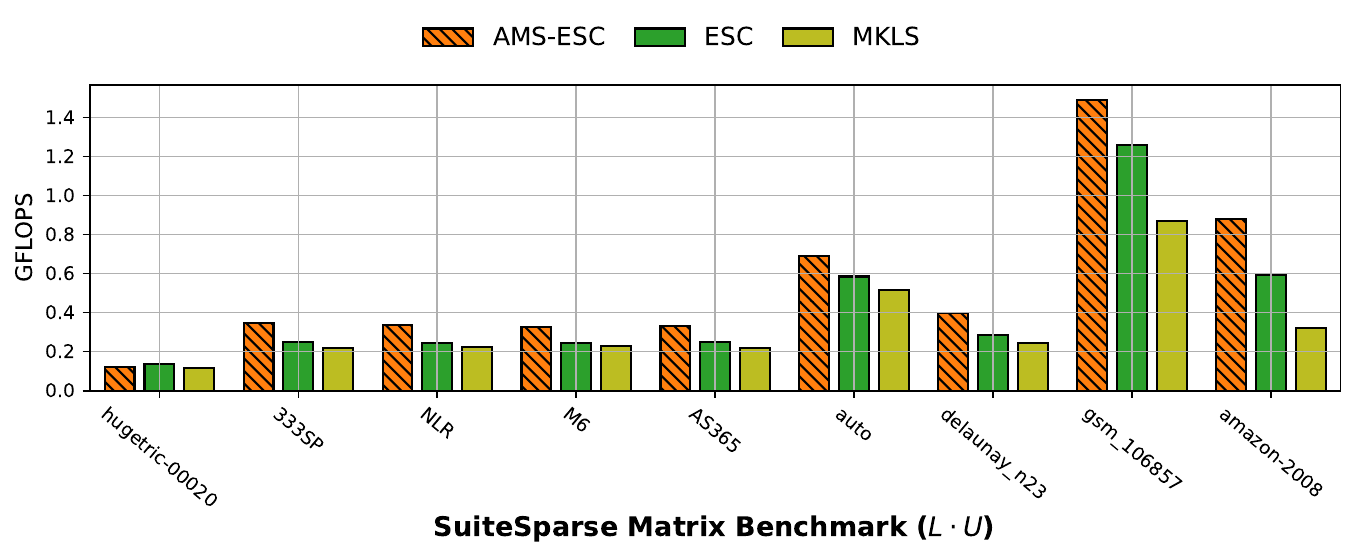}
	\caption{
		Triangle counting throughput on nine SuiteSparse matrices.
	}
	\Description{}
	\label{fig:tc}
\end{figure}

			Finally, we evaluate \myapp{} on triangle counting~\cite{DBLP:journals/pvldb/SundaramSPDAV0D15}, which counts the triangles in a graph.
			This operation supports clustering coefficient and transitivity calculations, as well as community detection and spam or anomaly detection in web and social networks.
			Triangle counting typically operates on an undirected graph, represented by a square, symmetric sparse adjacency matrix.
			In the formulation we evaluate, multiplying its strictly lower and upper triangular parts, $L \cdot U$, identifies paths of length two, also called wedges.
			The product is then masked with the adjacency matrix to retain only wedges whose endpoints are connected by an edge, forming a triangle.
			To apply this mask, each output row is merged with the corresponding adjacency row, requiring the entries within each output row to be sorted.

			Following the setup of Nagasaka et al.~\cite{DBLP:journals/pc/NagasakaMAB19}, we evaluate triangle counting on the nine square, structurally symmetric matrices in Table~\ref{tab:tamu}.
			For each matrix, we first reorder the vertices by increasing row nonzero count, applying the same permutation to rows and columns to preserve symmetry.
			We then split the reordered matrix into its strictly lower and upper triangular parts, $L$ and $U$, to form the SpGEMM operands.
			Since the subsequent masking step requires row-wise sorted output, we compare only sorted methods: \myapp{}-ESC, the ESC baseline, and the sorted Intel MKL variant (MKLS).

			Figure~\ref{fig:tc} reports the results.
			\myapp{}-ESC outperforms the ESC baseline on $8$ of the $9$ matrices, achieving a geometric-mean speedup of $1.27\times$ and a maximum speedup of $1.49\times$.
			It also outperforms MKLS on all $9$ matrices, with a geometric-mean speedup of $1.55\times$ and a maximum speedup of $2.76\times$.
			The largest gains over MKLS occur on the two most skewed matrices, amazon-2008 and gsm\_106857, with speedups of $2.76\times$ and $1.72\times$, respectively.
			The only loss to the ESC baseline occurs on hugetric-00020, the least skewed matrix in the benchmark (Gini $0.0003$), where there is little load imbalance to alleviate and insufficient computation to amortize the cost of $cf$-based morsel sorting.
			Together, these results demonstrate that \myapp{}'s benefits extend to real-world graph workloads requiring sorted output, with the greatest gains on skewed matrices.

	\subsection{Discussion}
	\label{subsec:disc}

		Our experimental analysis across sparse matrix multiplication, relational workloads, and graph analytics, on diverse datasets and hardware platforms, reveals several key findings that collectively validate the design rationale and practical effectiveness of \myapp{}.

		\textbf{Multi-dimensional skew is the primary performance bottleneck.}
		Across all matrix squaring evaluations (\S~\ref{subsec:perf}), we observe that the performance gap between \myapp{} and flop-count-based baselines widens as workload skew increases.
		On non-skewed ER matrices, \myapp{} performs comparably to the baselines, confirming that it introduces negligible overhead.
		On skewed G500 matrices, however, \myapp{} delivers substantial gains (up to $81.01\%$ for HASH), demonstrating that multi-dimensional skew, not algorithmic inefficiency, is the dominant factor limiting existing methods.
		That said, \myapp{} may occasionally underperform in highly sparse cases due to overhead from $cf$-based morsel sorting; however, this penalty diminishes rapidly as matrix density increases, and is outweighed by the substantial gains on the skewed workloads that are central to real-world applications.

		\textbf{Contention-aware scheduling is essential, not optional.}
		The dissection of \mysch{} and \mydaha{} (\S~\ref{subsec:dissect}) shows that simply achieving load balance is insufficient: methods that balance load but ignore cache contention (e.g., OpenMP dynamic and guided scheduling) still fall short of \mysch{}.
		Similarly, \mydaha{} offers a robust default that avoids the substantial degradation that either fixed policy incurs outside its sweet spot, confirming that the trade-off between computational throughput and cache pressure must be managed explicitly rather than left to a fixed policy.

		\textbf{Load balancing unlocks hardware potential.}
		The CXL evaluation (\S~\ref{subsec:cxl}) demonstrates that load imbalance can negate the benefits of bandwidth expansion, as straggler threads are disproportionately penalized by CXL's higher access latency.
		By addressing this imbalance, \myapp{} enables effective utilization of the augmented bandwidth, achieving throughput that exceeds the DRAM-only peak.
		This finding has broader implications. 
		As heterogeneous memory systems become more prevalent~\cite{DBLP:conf/edbt/JiH0HT24,DBLP:journals/fcsc/JiHZ24,weisgut2025cxl,DBLP:conf/asplos/LiuHWBNJNL25,DBLP:conf/asplos/LiuXBAL26}, contention-aware load balancing will be increasingly critical for realizing their performance potential.

		\textbf{The framework generalizes to other sparse kernels.}
		The SpMM and SpMV evaluation (\S~\ref{subsec:perf-spmmv}) shows that \myapp{}'s scheduling principles extend beyond SpGEMM to other sparse kernels with irregular sparsity patterns.
		The bandwidth measurements help explain these gains: on skewed workloads, the baseline uses only a fraction of the available DRAM bandwidth, while \myapp{} achieves higher bandwidth utilization.
		Since both kernels are memory-bound, contention-aware scheduling improves memory bandwidth utilization and thereby increases throughput.

		\textbf{SpGEMM-based methods are superior for relational query processing.}
		The join-aggregate and join-project evaluation (\S~\ref{subsec:perf-joinagg}) demonstrates the advantages of matrix-based methods, particularly when the intermediate join result is large.
		Traditional join-then-aggregate/project methods struggle with the resulting memory and processing costs, completing only the smallest workloads.
		SpGEMM-based methods avoid materializing this intermediate result and therefore run substantially faster.
		However, expressing these queries as sparse matrix products is only part of the solution.
		The performance limitations of prior matrix-based methods~\cite{deep2020fast,hu2020parallel,hu2025output,DBLP:conf/icde/SunKH23} highlight the need for efficient parallel execution.
		\myapp{} addresses this need through a load-balancing strategy that accounts for cache contention, translating the algorithmic benefits of the matrix formulation into consistent runtime gains.

		\textbf{The gains extend across graph-analytic workloads.}
		These benefits also extend to graph analytics, as shown by the evaluation in \S~\ref{subsec:perf-graph}.
		\myapp{} outperforms the strongest baseline across PathSim meta-path construction, multi-source BFS with non-sorted output, and triangle counting with sorted output.
		Together, these workloads demonstrate its effectiveness across different operand shapes, including tall-skinny and triangular matrices, and with both sorted and non-sorted output.
		Across these workloads, the gains are generally larger on more skewed inputs, reinforcing the practical value of contention-aware scheduling under skew.

\section{Related Work}
\label{sec:rw}

	SpGEMM has been extensively studied as a critical kernel in scientific computing and graph analytics~\cite{dalton2015optimizing,DBLP:conf/icde/ParkKJK023}, and more recently in relational query processing~\cite{deep2020fast,huang2022density,hu2020parallel,hu2025output,DBLP:conf/icde/SunKH23,DBLP:journals/pacmmod/KimR26}.
	These relational works, however, are largely confined to the algorithmic mapping from queries to sparse matrix products, and address neither the efficient parallel execution of the resulting kernel nor the load imbalance under skew that \myapp{} targets.
	Among them, SPALM~\cite{DBLP:journals/pacmmod/KimR26} is the most closely related to our work: it partitions the operands into small slices and selects a multiplication kernel for each slice pair based on local density and machine-calibrated thresholds.
	Its adaptivity focuses on kernel selection, leaving load balancing and cache contention outside its scope.
	\myapp{}, in contrast, adapts the size and execution order of parallel tasks to address both issues, making the two approaches complementary.

	Beneath every such mapping lies the SpGEMM kernel itself, whose optimization has been studied extensively.
	In this line of work, Gustavson's row-row formulation~\cite{DBLP:journals/toms/Gustavson78} has over the decades become the dominant approach due to its natural parallelizability, particularly when paired with the dense array accumulator known as SPA~\cite{gilbert1992sparse}.
	However, SPA has been shown to be inefficient and to incur high memory overhead, which prompted the development of more memory-efficient alternatives such as hash tables~\cite{richter2015seven,DBLP:journals/pc/NagasakaMAB19}, heaps~\cite{DBLP:conf/icppw/NagasakaMAB18}, and radix sort~\cite{dalton2015optimizing}.

	In addition to algorithmic advances, much attention has been given to the challenges of load balancing in parallel SpGEMM.
	Schemes based on flop counts~\cite{DBLP:journals/pc/NagasakaMAB19,DBLP:conf/icppw/NagasakaMAB18} have emerged as a dominant solution, along with other static strategies such as methods using blocks~\cite{DBLP:conf/icpp/BulucG08,DBLP:conf/ipps/BulucG08} or graph partitioning~\cite{DBLP:journals/siamsc/AkbudakA14}.
	While effective in certain scenarios, these approaches often fail to account for the multi-dimensional skew characteristics present in real-world SpGEMM workloads, resulting in imbalanced execution.

	As a more adaptive alternative, dynamic scheduling offers greater flexibility.
	For instance, Patwary et al.~\cite{DBLP:conf/supercomputer/PatwarySSPAV0PP15} proposed a dynamic task scheduling scheme that partitions work across both row and column dimensions.
	However, task partitioning across two dimensions introduces substantial synchronization overhead, which limits its practical applicability.
	Similarly, techniques leveraging work-stealing~\cite{DBLP:conf/ppopp/BlumofeJKLRZ95,yacsar2018fast,DBLP:conf/focs/Blumofe94} and cache-access modeling~\cite{DBLP:conf/icpp/GuoXLZZ24} have been explored, but either incur significant cache coherence overhead, or fail to account for complicated LLC sharing characteristics, making them less favorable than the flop-count-based approaches.
	Alongside these approaches, researchers have adapted dynamic variants of the classical longest-processing-time-first (LPT) algorithm~\cite{DBLP:journals/siamam/Graham69}, which schedules heavier tasks before lighter ones.
	However, none of these approaches account for on-chip cache contention caused by processing dense submatrices concurrently, an issue that \myapp{} explicitly addresses alongside multi-dimensional skew.

	Prior work has also explored task granularity in SpGEMM through one- and two-dimensional blocking~\cite{DBLP:conf/icpp/BulucG08,DBLP:conf/ipps/BulucG08}, sparse tiles~\cite{DBLP:conf/ppopp/NiuLJS0022,DBLP:conf/ppopp/HongSNSS19}, and adaptive panels~\cite{yao2024apspgemm}.
	These approaches balance spatial locality against the flexibility to distribute work across threads.
	A similar trade-off arises in data systems, where morsel-driven parallelism~\cite{DBLP:conf/sigmod/LeisBK014,DBLP:journals/pvldb/DursunBCSG19} distributes fixed-size groups of \emph{input} tuples across workers.
	\myapp{} adopts the morsel abstraction but varies the input size to maintain a roughly fixed \emph{output} volume per morsel.
	It also orders morsel execution to reduce shared-cache contention, which conventional morsel-driven parallelism does not explicitly address.
	The approaches are therefore complementary, and \myapp{}'s techniques could benefit morsel-driven engines by using partition-size estimates to guide dispatch order and SMT activation for hash-based operators under skew.

\section{Conclusion}
\label{sec:concl}
	This paper presents \myappfull{} (\myapp{}), a morsel-wise dynamic scheduling framework that addresses two co-existing challenges in SpGEMM, namely \emph{multi-dimensional skew} and \emph{on-chip cache contention}.
	\myapp{} balances skewed workloads without aggravating cache contention through \myschfull{} (\mysch{}), judiciously activates SMT via \mydahafull{} (\mydaha{}), and adaptively selects the per-morsel accumulator with \myaccumfull{} (\myaccum{}).
	Evaluations across matrix squaring, relational queries, graph analytics, and other sparse kernels show that \myapp{} significantly outperforms state-of-the-art methods on skewed workloads while maintaining comparable performance on non-skewed workloads.
	It also improves memory bandwidth utilization, with these benefits observed on both conventional and CXL-expanded memory systems.
	Looking ahead, extending the morsel-wise scheduling framework to other SpGEMM formulations, such as outer-product and inner-product methods, could further improve performance on highly sparse matrices.

\bibliographystyle{ACM-Reference-Format}
\bibliography{ref}

\end{document}